\documentclass[twocolumn]{aastex63}
\usepackage{amsmath}

\usepackage{lineno}

\newcommand{\be}{\begin{eqnarray}}
\newcommand{\ee}{\end{eqnarray}}

\graphicspath{{./}{figures/}}

\newcommand\js{\bgroup\markoverwith
{\textcolor[rgb]{0.45, .0, .85}{\rule[.5ex]{8pt}{1.5pt}}}\ULon}
\newcommand\kqs{\bgroup\markoverwith
{\textcolor[rgb]{0.85, .0, .45}{\rule[.5ex]{8pt}{1.5pt}}}\ULon}

\submitjournal{ApJ} 

\shorttitle{Binary formation in AGN disks}
\shortauthors{Qian, Li, and Lai}

\graphicspath{{./}{figures/}}

\begin{document}

\title{Dynamical Friction Models for Black-Hole Binary Formation in AGN Disks
}

\correspondingauthor{Kecheng Qian}
\email{kq38@cornell.edu}

\author[0009-0000-3886-8014]{Kecheng Qian}
\affiliation{Center for Astrophysics and Planetary Science, Department of Astronomy, Cornell University, Ithaca, NY 14853, USA}

\author[0000-0001-5550-7421]{Jiaru Li}
\affiliation{Center for Astrophysics and Planetary Science, Department of Astronomy, Cornell University, Ithaca, NY 14853, USA}
\affiliation{Center for Interdisciplinary Exploration and Research in Astrophysics (CIERA), Northwestern University, 1800 Sherman Ave, Evanston, IL 60201, USA}

\author[0000-0002-1934-6250]{Dong Lai}
\affiliation{Center for Astrophysics and Planetary Science, Department of Astronomy, Cornell University, Ithaca, NY 14853, USA}
\affiliation{Tsung-Dao Lee Institute, Shanghai Jiao-Tong University, Shanghai, 520 Shengrong Road, 201210, China}

\begin{abstract}
Stellar-mass black holes (sBH) embedded in gaseous disks of active galactic nuclei (AGN) can be important sources of detectable gravitational radiation for LIGO/Virgo when they form binaries and coalesce due to orbital decay. 
In this paper, we study the effect of gas dynamical friction (DF) on the formation of BH binaries in AGN disks using $N$-body simulations. We employ two simplified models of DF, with the force on the BH depending on $\Delta {\bf v}$, the velocity of the sBH relative to the background Keplerian gas. 
We integrate the motion of two sBH initially on circular orbits around the central supermassive black hole (SMBH), and evaluate the probability of binary formation under various conditions. 
We find that both models of DF (with different dependence of the frictional coefficient on $|\Delta{\bf v}|$) can foster the formation of binaries when the effective friction timescale $\tau$ satisfies $\Omega_{\rm K}\tau\lesssim 20-30$ (where $\Omega_{\rm K}$ is the Keplerian frequency around the SMBH): prograde binaries are formed when the DF is stronger (smaller $\tau$), while retrograde binaries dominate when the DF is weaker (larger $\tau$). 
We determine the distribution of both prograde and retrograde binaries as a function of initial orbital separation and the DF strength. 
Using our models of DF, we show that for a given sBH number density in the AGN disk, the formation rate of sBH binaries increases with decreasing $\tau$ and can reach a moderate value with a sufficiently strong DF.
\end{abstract}



\section{Introduction}

Mergers of stellar-mass binary black holes (BBHs) are the most common gravitational waves (GW) sources detected by LIGO/Virgo \citep{LIGO2019ApJL,LIGO2021ApJL05,LIGO2021arXiv08,LIGO2021arXiv11}. 
An interesting venue for these merger events is the accretion disks around supermassive black holes (SMBHs) in active galactic nuclei (AGN) \citep[e.g.,][]{McKernan2012MNRAS, McKernan2014MNRAS, Bartos2017ApJ, Stone2017MNRAS, Secunda2019ApJ, Secunda2020ApJ, Yang2019ApJ, Yang2019PRL, Grobner2020aap, Ishibashi2020aap, Tagawa2020ApJ07, Tagawa2020ApJ08, Ford2021arxiv},
where mergers of stellar-mass black-holes (sBHs) have been argued to exhibit distinct observable distributions of mass, spin, and eccentricity \citep[e.g.,][]{McKernan2018ApJ,Yang2019ApJ,Gerosa2021NatAstro,Tagawa2021MNRAS,Wang2021ApJ,LJR2022ApJ,Samsing2022Natur}, and may also produce electromagnetic counterparts that are associated with the GW signals \citep[e.g.,][]{Stone2017MNRAS,McKernan2019ApJL,Graham2020PRL,Graham2023ApJ,Palmese2021ApJ}. However, large uncertainties remain concerning the occurrence rate of such mergers and the dynamical roles of gas disks in the formation and evolution of the embedded sBH binaries.

Several mechanisms have been proposed for facilitating the mergers of BBHs in AGN disks. 
For example, the dynamical evolution of BBHs can be affected by the gravity from the highly perturbed gas around them \citep[e.g.,][]{Baruteau2011ApJ,Stone2017MNRAS}; 
recent high-resolution hydrodynamics simulations have shown that, for some AGN disk parameters, such BBH-disk interaction may indeed help the BBHs contract in orbit (and possibly merge) \citep{LiYP2021ApJ,LiYP2022ApJL,Dempsey2022ApJ,LiRX2022MNRAS,LiRX2023MNRAS}. 
BBHs embedded in an AGN disk may also be hardened and driven to mergers by a series of scatterings with other single BHs \citep[e.g.,][]{Stone2017MNRAS, Leigh2018MNRAS, Samsing2022Natur}, especially when the mutual orbital inclinations between the BHs are small.

These mechanisms, however, all assume pre-existing binaries.
The origin of sBH binaries in AGN disks remains an open question. 
While some binaries may be formed in nuclear star clusters then captured into the AGN disk \citep{Bartos2017ApJ}, it has been suggested that many binaries may form within the AGN disk through close encounters between embedded single BHs \citep[e.g.,][]{Stone2017MNRAS,Fabj2020MNRAS}. 
Since AGN disks may have non-monotonic radial density profiles \citep{Sirko2003MNRAS,Thompson2005ApJ,Dittmann2020MNRAS,Gilbaum2022ApJ}, single BHs embedded in a disk may naturally approach each other due to differential migrations or migration traps \citep{Bellovary2016ApJL,Secunda2019ApJ,Secunda2020ApJ}. 
The resulting tightly-packed BHs orbits can be dynamically unstable, leading to frequent close encounters and opportunities for binary formation.

A few recent studies have investigated the dynamics and the probability of forming tightly bound BBHs through close encounters.
Assuming negligible interactions between the BHs and the gaseous disk, \cite{LJR2022ApJ} showed by a suite of long-term $N$-body simulations that BH-BH encounters almost always lead to weakly bound and short-lived binaries; 
only in the rare event of very close encounters, which may occur over longer timescales, the two BHs can be captured into tightly bound binaries via GW emission. 
Such binaries are formed with very high eccentricities, and remain eccentric when entering the LIGO band. 
Combining the simulation results with a theoretical scaling analysis, they obtained the probability of binary formation via this ``GW-bremsstrahlung'' mechanism (see \cite{rom2023formation} for an analytical study). 
Binary formation without gas was also studied by \cite{Boekholt2023MNRAS} using a phase space fractal structure analysis.

\cite{Tagawa2020ApJ07} estimated that most merging BBHs in AGN disks are formed in close encounters through the dissipation of kinetic energy in the disk gas, although their results are based on one-dimensional $N$-body simulations and rough semi-analytical prescriptions (i.e., with simplified binary evolutions and BH-disk interactions).
\cite{LJR2023ApJL} used high-resolution global hydrodynamics simulations to study the close encounters between two single sBHs in AGN disks.
They showed that when the disk density is sufficiently high, bound BH binaries can be formed by the collision of circum-single disks.
They also found that the resulting binaries tend to have compact semi-major axes, large eccentricities, and retrograde rotations.
A similar investigation using smoothed-particle hydrodynamics simulations was done by \cite{Rowan2023MNRAS}, who found that prograde binaries can also be formed. 
Both \cite{LJR2023ApJL} and \cite{Rowan2023MNRAS} have only surveyed limited parameter spaces because their simulations are very expensive. 
\cite{DeLaurentiis2023MNRAS} studied the BH-BH close encounters in AGN disks using a large number of $N$-body simulations that model the gas effect with a simple gas dynamical friction \citep[GDF,][]{Ostriker1999ApJ}. 
They characterized the properties of the resulting binaries as functions of the impact parameter of the close encounters, and explored the dependence of BBH formation on various parameters such as BH mass, disk temperature, and disk density.
Recently, \cite{Whitehead2023arXiv} carried out extensive hydrodynamical simulations in shearing box, and determined the condition of binary formation as a function of the impact parameter of the two sBHs and the gas density of the background AGN disk.

Although previous works have found that BH-BH encounters can form both prograde and retrograde binaries, a complete picture of the conditions and the rates of their formation is still missing.
Prograde and retrograde binaries (relative to the original AGN disk) can affect the spin-orbit misalignment angles of the merging BBHs.
It has also been recognized that, compared to the prograde ones, retrograde binaries may inspiral at faster rates and can grow their masses more rapidly through gas accretions \citep{LiYP2021ApJ,LiRX2022MNRAS}.
Hence, understanding the conditions and rates for the formation of prograde and retrograde binaries is important for assessing the AGN channel of BBH mergers.

In this paper, we study the gas-assisted formation of prograde and retrograde BH binaries in an AGN disk through $N$-body simulations.
Using two different gas force models, we systematically investigate the binary formation conditions
as functions of the BHs' initial orbital separation, masses, and the strength of the gas forces.
Our simulations employ a similar setup as in some previous works \citep[e.g.,][]{LJR2022ApJ, DeLaurentiis2023MNRAS}.
However, we focus on the results that can be scaled to various parameters/conditions, and our calculations reveal the different pathways and probabilities to form binaries with different rotational orientations.

The rest of this paper is organized as follows. 
In Section~\ref{sec:method}, we describe the setup of our $N$-body simulations, and the two different gas force models, including the linear frictional force and the GDF.
Sections~\ref{sec:M1} and~\ref{sec:M2} show the results of the $N$-body simulations with the linear frictional force and the GDF, respectively.
We use a local shearing-box model to analyze our simulation results in Section~\ref{sec:shearing_box}.
We include GW emission in Section \ref{sec:GW-binary} and examine the relative roles of gas friction and GW-bremsstrahlung.
Then, based on these results, Section~\ref{sec:FR} estimates the formation rates of prograde and retrograde binaries in AGN disks (given the number density of sBHs).
Section~\ref{sec:summary} summarizes our findings.

\section{Method}
\label{sec:method}

\subsection{Simulation Setup and Initial Conditions}
\label{sec:method_ic}

We consider a system consisting of a central SMBH (mass $M$), and two sBHs ($m_1$ and $m_2$) that orbit around $M$ on initially separate, nearly circular orbits.
The sBHs are embedded in a gaseous accretion disk around the SMBH and are subjected to gas drags.
We investigate how the two sBHs dynamically form a bound binary with the assistance of the gas.

The two sBHs are given masses of $m_1=10^{-6}M, m_2=0.5\times 10^{-6}M$ for the most part of this paper.
They are initially placed on co-planar circular Keplerian orbits around the SMBH with orbital radii $a_1$ and $a_2$. 
We set the initial orbital separation of $m_1$ and $m_2$ using the dimensionless parameter $K$, defined as
\begin{equation}
    K\equiv \frac{a_2-a_1}{R_{\rm H}},
\end{equation}
where
\begin{equation}
    \label{hillradius}
    R_{\rm H}=\frac{a_1+a_2}{2}\left( \frac{m_1+m_2}{3M} \right)^{1/3}\simeq a_1\left( \frac{m_{12}}{3M}\right)^{1/3}
\end{equation}
is the mutual Hill radius of the two sBHs with $m_{\rm 12}=m_1+m_2$.
We also specify the initial azimuthal separation of the sBHs by the angle $\Delta \phi_0 =\phi_2-\phi_1$ where $\phi_1,\phi_2$ are the initial orbital phases. 
We choose the values of $K$ from the range $[1.0, 2.5]$ and $\Delta \phi_0$ from $[10R_{\rm H}/a_1, 2\pi-10R_{\rm H}/a_1]$. 
These values allow $m_1$ and $m_2$ to be well-separated ($\gg R_{\rm H}$) initially and become much closer ($\lesssim R_{\rm H}$) at later time.
The sBHs are then expected to experience a close encounter and can possibly form a binary.
Note that for $K$ outside this range, the orbits of the sBHs are either weakly perturbed (for $K \gtrsim 2.5$) or experience strong deflection on a horseshoe trajectory (for $K\lesssim 1$) (see \citealt[e.g.][]{Petit1986Icar}), and binary formation is nearly impossible, except for a few cases enabled by multiple close encounters (see Section \ref{sec:M1}).

\subsection{Running the Simulation}

We simulate the orbital evolution of the ``SMBH+2sBHs'' systems using the $N$-body software \texttt{REBOUND} \citep{Rein2012aap} and the \texttt{IAS15} integrator \citep{Rein2015MNRAS}.
The calculations are repeated for many values of $K$ and $\Delta \phi_0$.
Each initial condition is run for $150P_{\rm K}$, where
\begin{equation}
    \label{eq:Omega_Kep}
    P_{\rm K}=2\pi \Omega_{\rm K}^{-1} \simeq P_{1,2}=2\pi \sqrt{\frac{a_{1,2}^3}{GM}}
\end{equation}
is the initial orbital period of the sBH around the SMBH and $\Omega_{\rm K}\simeq \sqrt{GM/a_{1,2}^3}$ is the initial Keplerian angular velocity. Note that the synodic period of the two sBHs (neglecting mutual interaction) is ($2a_1/3KR_{\rm H}$)$P_{\rm K}\simeq (80/K)P_{\rm K}$ for our canonical BH masses.
Hence, the time limit of $150P_{\rm K}$ allows the sBHs to experience about two close encounters for most cases \citep[see Section~\ref{sec:M1_small_K} and also see Figures~12 and 13 of][]{LJR2022ApJ}.

The gas effects are implemented in our $N$-body simulations as additional forces on the sBHs.
We apply two models of gas forces, as described in the next subsection.

\subsection{Models for gas-BH interaction}
In this work, we consider two models for the gas drags experienced by the sBHs in the Keplerian AGN disk.

\subsubsection{Model 1}
The first model applies a linear frictional force. The force per unit mass is given by
\begin{equation}
\label{eq:gas-force-1}
    {\bf a}=-\frac{{\bf v}-{\bf v}_{\rm K}}{\tau},
\end{equation} 
where $\bf{v}$ is the velocity of the sBH and ${\bf v}_{\rm K}$ is the local Keplerian velocity of the disk around the SMBH.
The strength of this friction is determined by the timescale $\tau$.

\begin{figure}
    \centering
    \includegraphics[width=0.45\textwidth]{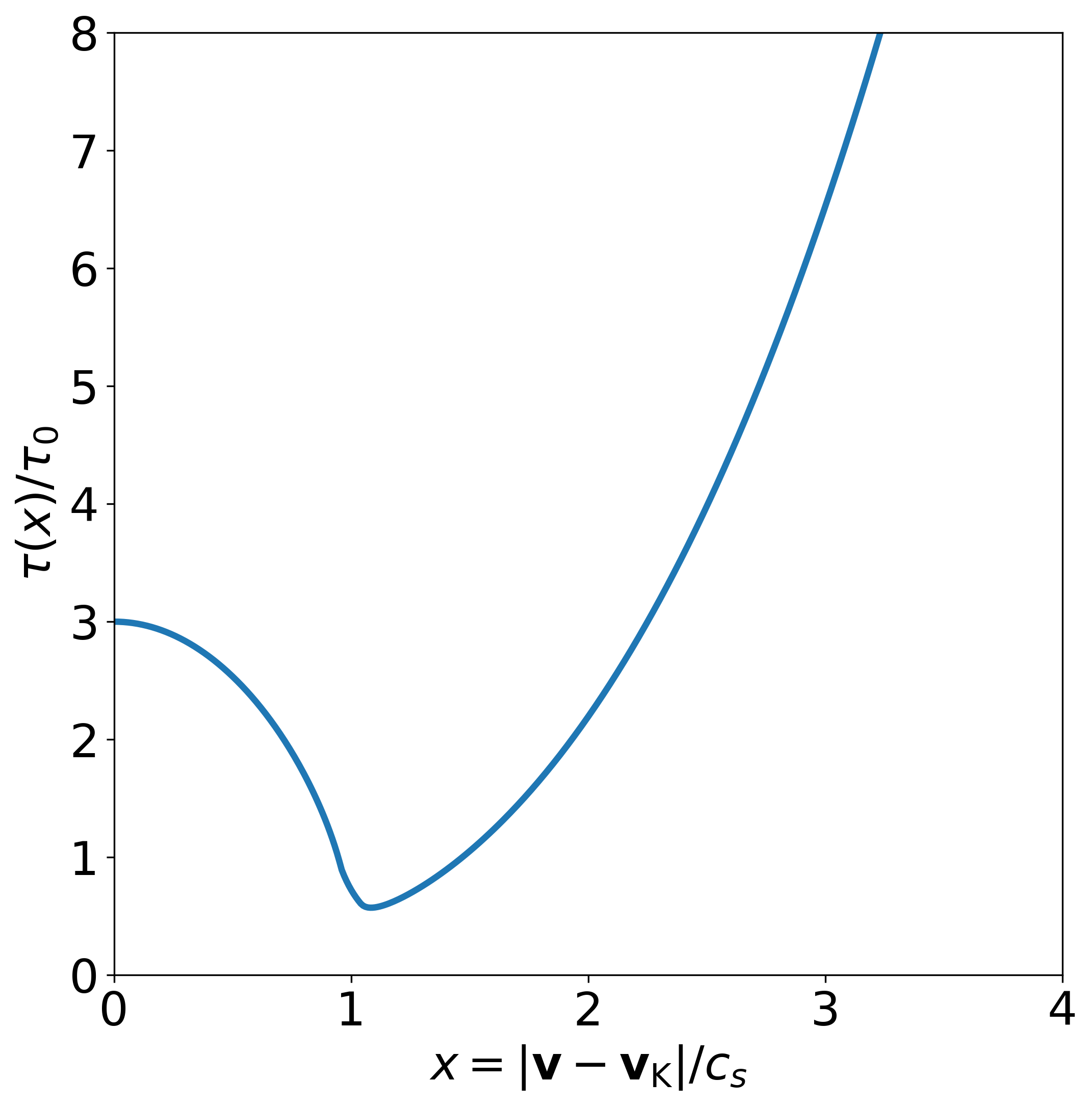}
    \caption{
    Dissipation timescale $\tau(x)$ used in our Model 2 simulations, where $x\equiv|{\bf v}-{\bf v}_{\rm K}|/c_s$ is the Mach number of the relative velocity between the sBH and the Keplerian gaseous disk.
    The constant $\tau_0$ is given by Equation~\eqref{eq:tau0}.} 
    \label{x^3/f(x)}
\end{figure}

\subsubsection{Model 2}
\label{sec:method-gas-gdf}

The second model is the canonical gas dynamical friction \citep[GDF,][]{Ostriker1999ApJ}. We express the GDF as
\begin{align}
\label{eq:gas-force-2}
    {\bf a}=-\frac{{\bf v}-{\bf v}_{\rm K}}{\tau(x)},
\end{align}
which is similar to Model 1 except that the friction timescale 
\begin{align}
    \tau(x)=\tau_0\frac{x^3}{f(x)}
\end{align}
is not a constant.
The variable $x \equiv |{\bf v}-{\bf v}_K|/c_s$ is the Mach number of the relative velocity between the sBH and the background gas, while the coefficient
\begin{equation}
    \label{eq:tau0}
    \tau_0=\frac{c_s^3}{4\pi G^2 \rho m}
\end{equation}
is a characteristic damping timescale determined by the mass of the sBH ($m$), the gas density $\rho$, and the sound speed $c_s$. 
For simplicity, we assume a constant $\tau_0$ (but $m$-dependent) for each sBH and a constant disk aspect ratio $h=c_s/|{\bf v}_K|$ for each of our simulations (i.e. $\tau_0$ is independent of the sBH position).

The function $f(x)$ is defined piece-wisely in the subsonic and supersonic regimes as \citep{Ostriker1999ApJ}
\begin{equation}
    \label{f(x)}
    f(x)=\begin{cases}
      0.5\ln(\frac{1+x}{1-x})-x,&0<x<1,\\
      0.5\ln(x^2-1)+3.1,\ \ &x>1,
    \end{cases}
\end{equation}
In practice, to achieve fast convergence of numerical integrations, we adopt two approximations.
First, when $x\to0$, we note that both $f(x)$ and $x^3$ approach zero but $x^3/f(x)\to 3$. Thus we set $\tau(0)=3\tau_0$ and adopt a linear interpolation to calculate $\tau(x)$ between $x=0$ and $x=0.01$.
Second, to avoid the discontinuity in $f(x)$ at $x=1$, we linearly interpolate $\tau(x)$ between $x=1-\delta$ and $x=1+\delta$ based on the exact values of $\tau$ at $x=1-\delta$ and at $x=1+\delta$. To ensure the smoothness of $\tau(x)$, we choose $\delta=0.04$. 
After the two treatments, our smoothed dissipation timescale $\tau(x)$ is shown in Figure~\ref{x^3/f(x)}.

\section{N-Body Simulation Results: Binary Formation with Gas Friction Model 1}
\label{sec:M1}

\begin{figure*}
    \centering
    \includegraphics[width=1\textwidth]{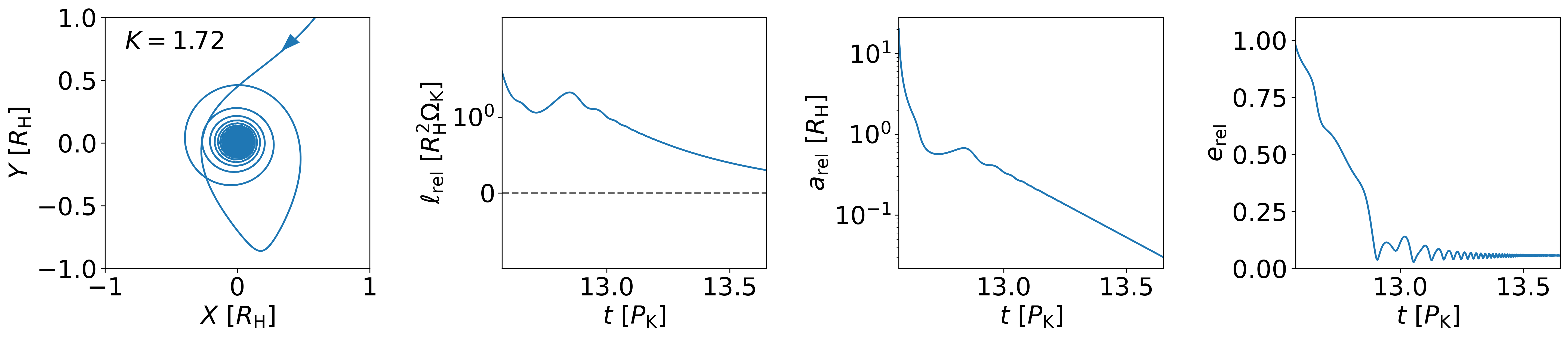}
    \vfill
    \includegraphics[width=1\textwidth]{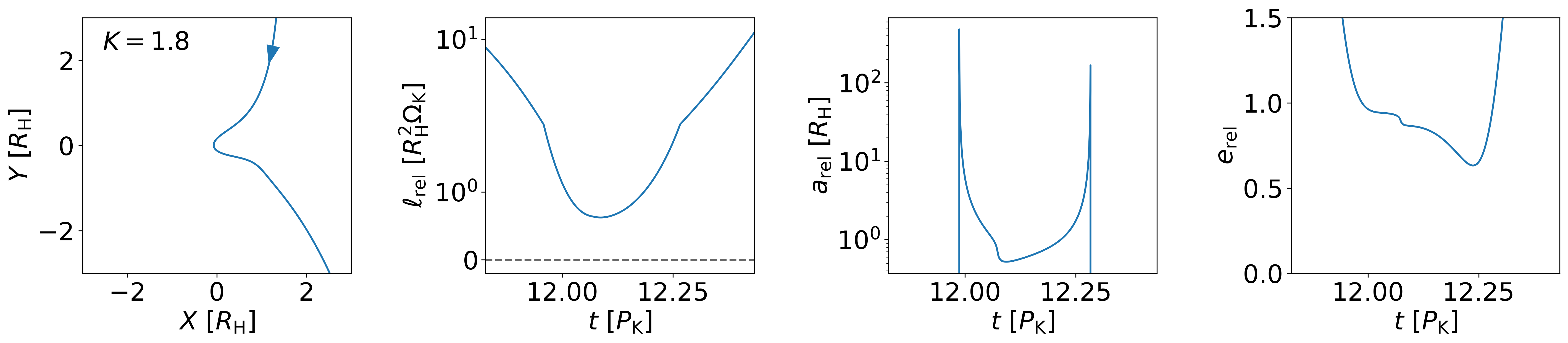}
    \vfill
    \includegraphics[width=1\textwidth]{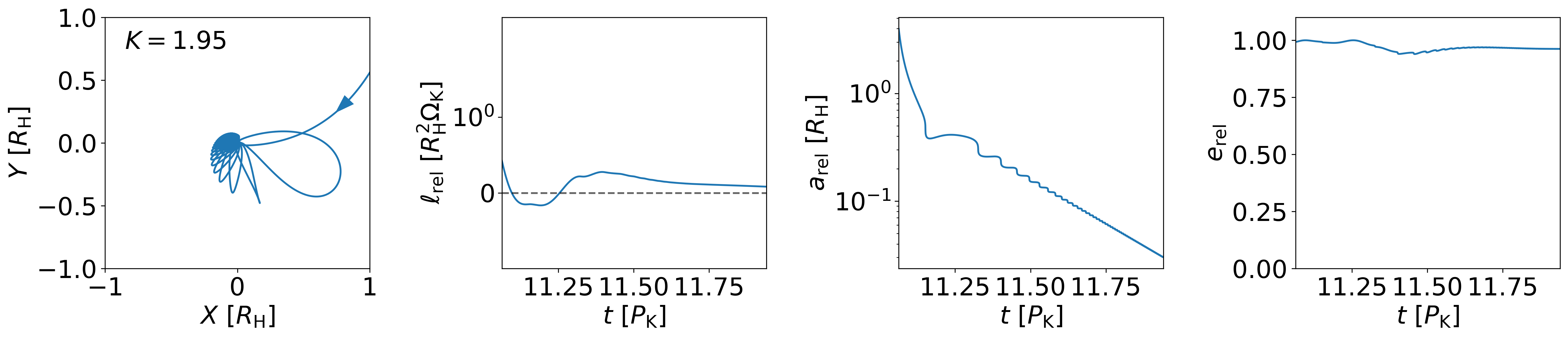} 
    \vfill
    \includegraphics[width=1\textwidth]{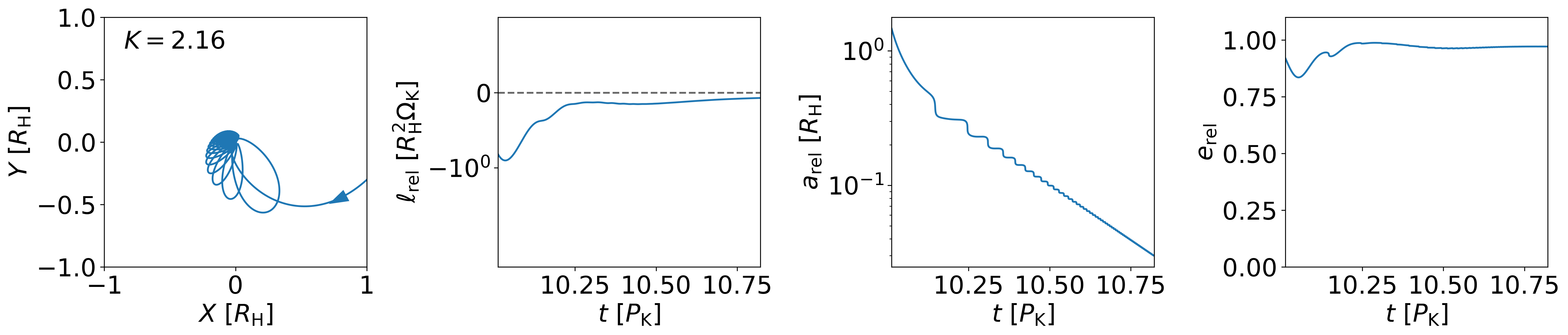} 
    \vfill
    \includegraphics[width=1\textwidth]{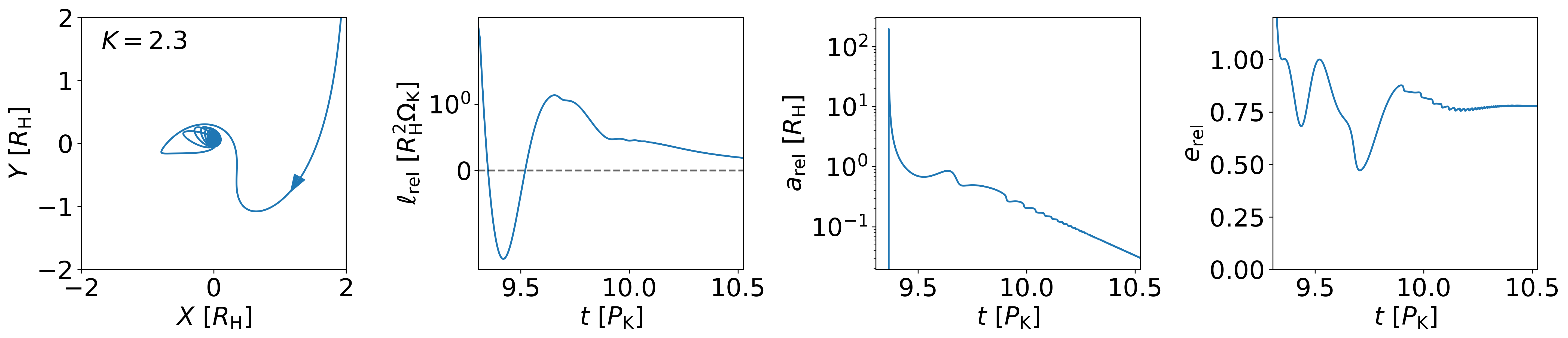} 
    \caption{Selected results of the $N$-body simulations with the linear frictional force (Model 1, equation~\ref{eq:gas-force-1}).
    Each row represents one simulation: from left to right, the panels show the relative trajectory of the sBHs in the co-rotating frame of $m_1$ (with $X$ and $Y$ being the radial and azimuthal component of ${\bf r}_2-{\bf r}_1$), the specific relative angular momentum of the sBHs $\ell_{\rm rel}=\bf{l}_{\rm rel}\cdot \bf{z}$ (equation~\ref{eq:l_rel}), and the osculating orbital elements $a_{\rm rel}$ (equation~\ref{eq:a_bin}) and $e_{\rm rel}=\sqrt{1-\ell_{\rm rel}^2/(G m_{12}a_{\rm rel})}$.
    The simulations adopt $1/\tau=0.3\Omega_{\rm K}^{-1}$, $m_1=10^{-6}M$, $m_2=0.5\times 10^{-6}M$, $\Delta \phi_0=\pi/2$, and the values of the initial $K$ are shown in the left panels. Prograde binaries (with $\ell_{\rm rel}>0$) are formed in rows 1,3 and 5, retrograde binary (with $\ell_{\rm rel}<0$) in row 4, no binary formation is observed in row 2.
    }
    \label{fig:model1_trajectory}
\end{figure*}

In this section, we run $N$-body simulations for the ``SMBH+2sBHs'' systems with the gas frictional force Model 1.

Figure~\ref{fig:model1_trajectory} shows the results of sBHs in five example runs.
Each row represents one simulation: the left-most panel shows the relative trajectory of the sBHs; the second-left, second-right, and right-most panels depict the specific angular momentum, semi-major axis, and eccentricity of the sBHs' relative orbit, respectively.

These examples show two apparently distinct outcomes: the sBHs either form (rows 1 and 3 to 5) or do not form bound binaries (row 2). 
Among the formed cases, both prograde (rows 1, 3, and 5) and retrograde binaries (row 4) are observed. 

Clearly, the outcomes of the close encounters may depend on the initial conditions ($K$, $\Delta \phi_0$), the masses of sBHs $m_{1,2}$, and the gas damping timescale $\tau$. Therefore, we conduct parameter studies to investigate the effect of varying these parameters.

\subsection{Criteria for binary formation}
\label{sec:criteria}

Before showing the results of our parameter study, we first discuss our prescription to identify binary formation, and determine the sense of rotation (prograde/retrograde) of the formed binaries. 
This is needed to efficiently categorize the simulation results in the parameter survey.

Given our choice of 150$P_{\rm K}$ as the integration time, the two BHs would have sufficient interaction time after their first close encounter. We consider the binaries have formed if the following two conditions are \textit{both} met:

(i) The separation between the two BHs remains smaller than $R_{\rm H}$ until the end of integration, after reaching $R_{\rm H}$ at some time.

(ii) The relative energy of $m_1, m_2$ becomes negative at some time and remains negative until the end of integration, i.e.
\begin{equation}
    \label{eq:Erel_inertial}
    E_{\rm rel}=\frac{1}{2}\mu |{\bf v}_1-{\bf v}_2|^2-\frac{Gm_1m_2}{|{\bf r}_1-{\bf r}_2|}<0,
\end{equation}
where $\mu=m_1m_2/m_{12}$ is the reduced mass, and ${\bf r}_{1,2}$ and ${\bf v}_{1,2}$ are the position and velocity vectors of $m_{1,2}$. 

In practice, the orbital integration becomes more time-consuming, when the separation $|{\bf r}_1-{\bf r}_2|$ gets too small.
To avoid such a situation, we use a third condition to terminate the integration and directly conclude that the binary has formed once the condition is satisfied: 

(iii) The relative semi-major axis of $m_1, m_2$ is positive and less than $0.1R_{\rm H}$, i.e. 
\begin{equation}
\label{eq:a_bin}
    a_{\rm rel}=-\frac{Gm_1 m_2}{2 E_{\rm rel}}<0.1R_{\rm H}
\end{equation}
This condition implies that the relative apocenter separation of the BHs is less than $0.2R_{\rm H}$. 

We have verified that the above conditions provide good criteria for discerning whether the two BHs form stable binaries in our numerical integrations.

We determine the orientation of the binaries by computing the specific relative angular momentum of the sBHs,
\begin{equation}
    \label{eq:l_rel}
    {\bf l}_{\rm rel}=({\bf r}_2-{\bf r}_1)\times ({\bf v}_2-{\bf v}_1).
\end{equation}
With ${\bf z}$ being the unit vector along the sBHs' initial orbital angular momentum around the SMBH, a binary is considered prograde if $\ell_{\rm rel}={\bf l}_{\rm rel}\cdot {\bf \hat z}>0$ (i.e., the binary's relative orbit is in the same direction as its center-of-mass orbit about the SMBH) and retrograde if $\ell_{\rm rel}={\bf l}_{\rm rel}\cdot {\bf \hat z}<0$. 

In some simulations, the sign of $\ell_{\rm rel}$ 
might change a few times before stabilizing (see the second-left panel of the bottom row of Figure~\ref{fig:model1_trajectory}.) We determine the orientation of a binary based on the direction of ${\bf l}_{\rm rel}$ at the end of the integration time.

\subsection{Parameter Study Results}
\label{sec:M1_param}

\subsubsection{Dependence on $K$ and $\tau$}
\label{sec:M1_param_K}
\begin{figure}
\centering
    \includegraphics[width=0.9\columnwidth]{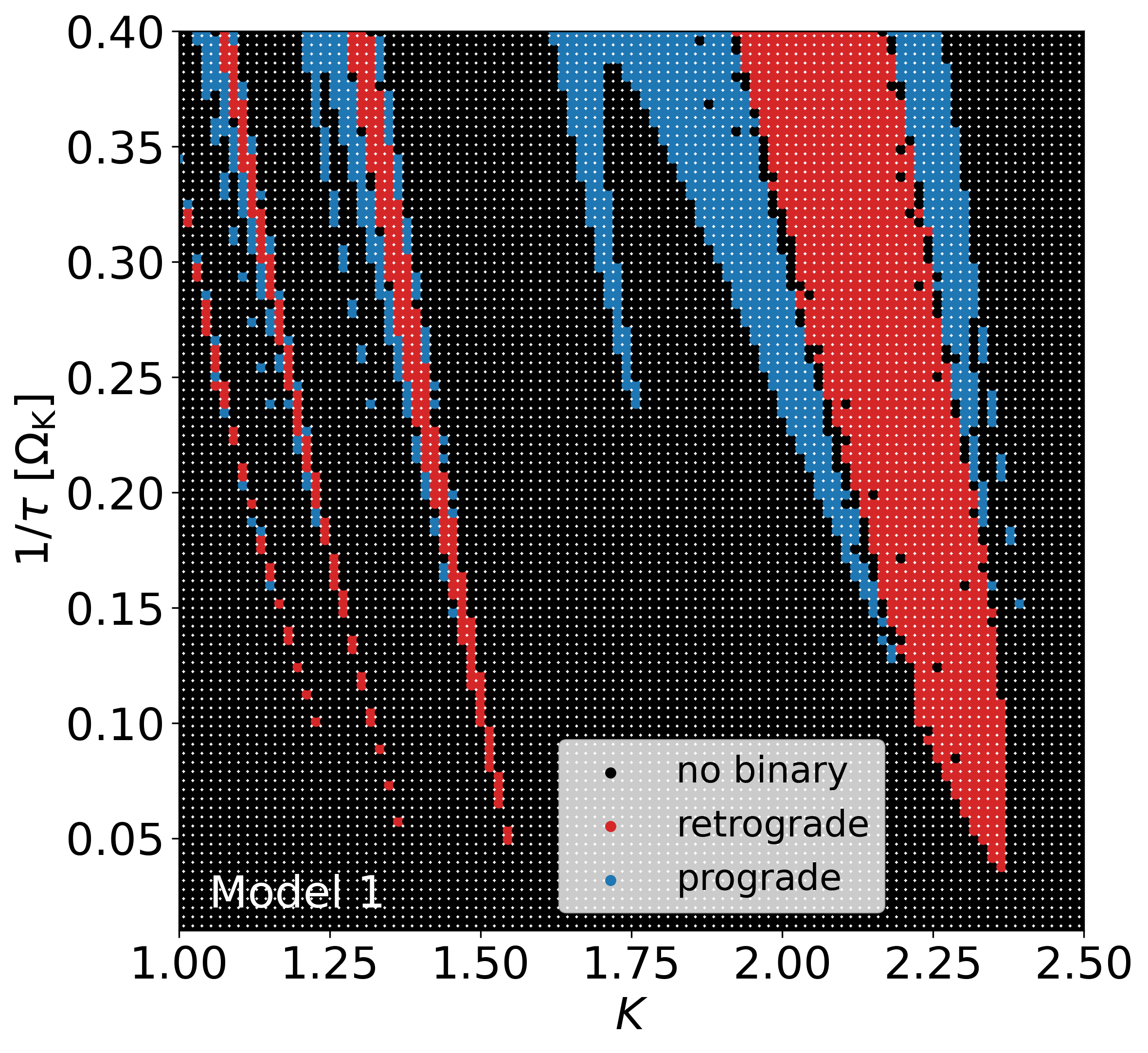}
    \caption{
    Outcomes of $N$-body simulations with different initial orbital separations $K$ and frictional damping rate $\tau^{-1}$ (in units of $\Omega_{\rm K}$, the initial orbital frequency of $m_1$).
    All simulations adopt $m_1=10^{-6}M$, $m_2=0.5\times 10^{-6}M$, $\Delta \phi_0=\pi/2$, and the Model 1 gas force (equation~\ref{eq:gas-force-1}).
    Each dot represents an integration. 
    The black, blue, and red dots represent the distribution of simulations that do not form binaries, form prograde binaries, and form retrograde binaries, respectively.
    }
    \label{fig:reboundmodel1}
\end{figure}

We run a suite of simulations with the initial $K$ in the range $[1,2.5]$, $1/\tau$ in the range $[0,0.4\Omega_{\rm K}]$, and the initial $\Delta \phi_0 = \pi/2$.
Figure~\ref{fig:reboundmodel1} shows the result:
each dot represents one integration with a different combination of $K$ and $\tau$; a black dot means binary is not formed, a blue dot means prograde binary is formed, and a red dot means retrograde binary is formed.

This result suggests that the formation outcome strongly depends on the initial $K$ and the friction timescale $\tau$.
BH binaries are only formed when $1/\tau$ exceeds a critical value, given by
\begin{equation}
    \label{eq:tau_crit}
    \tau^{-1}_{\rm crit}\simeq0.03\Omega_{\rm K} 
\end{equation}
All binaries are retrograde when $1/\tau$ is approximately between $0.03\Omega_{\rm K}$ and $0.13\Omega_{\rm K}$. 
Prograde binaries appear when $1/\tau\gtrsim0.13\Omega_{\rm K}$.
Most binary formations are formed in systems with the initial $K \in [1.7,2.4]$. 
For initial $K\lesssim1.7$, binaries are only formed in a few systems with some special combination of $K$ and $\tau$ (see Section~\ref{sec:M1_small_K}).

\subsubsection{Dependence on $\Delta \phi_0$}

\begin{figure}
    \centering
    \includegraphics[width=0.9\columnwidth]{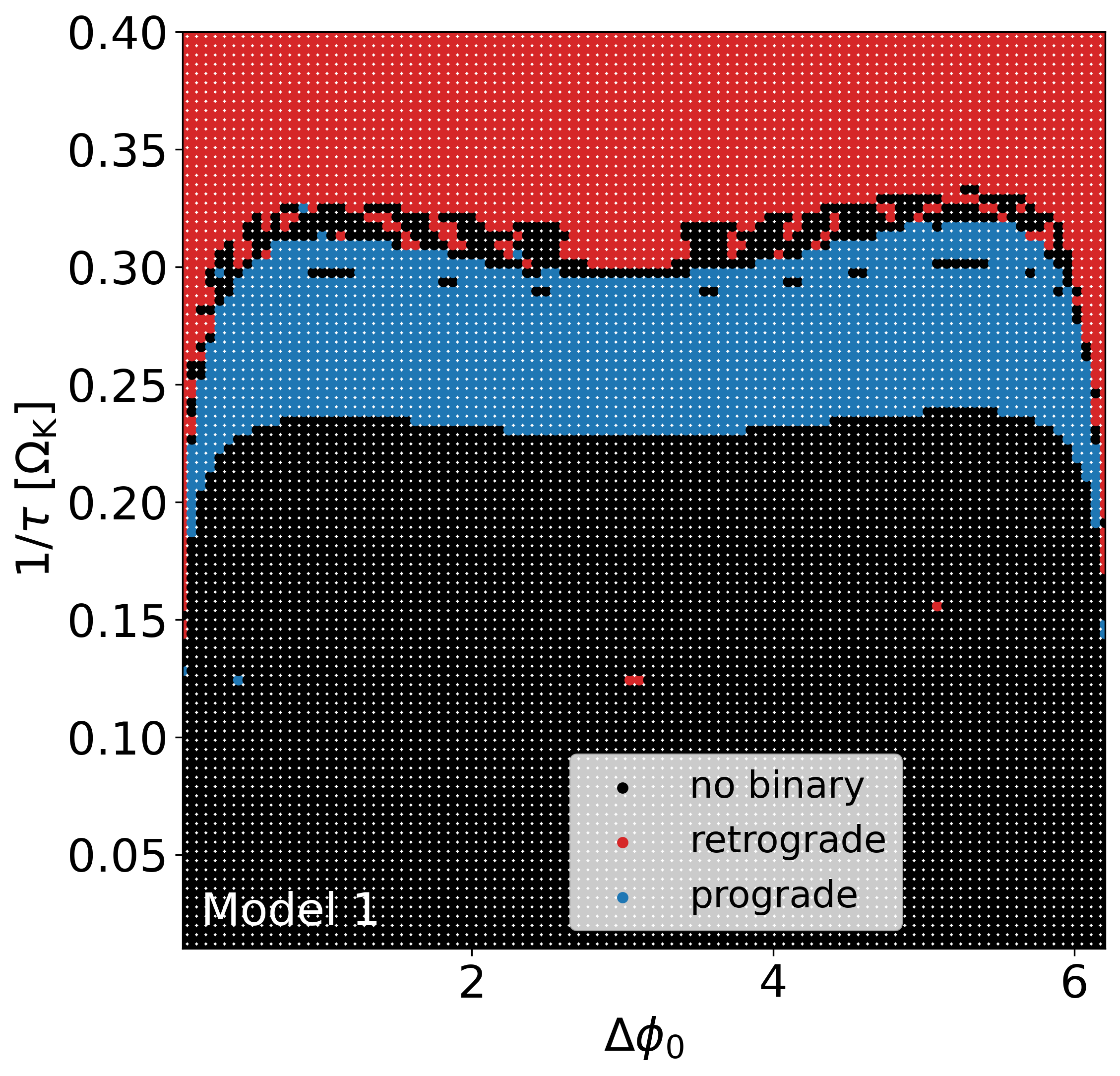}
    \caption{
    Same as Figure~\ref{fig:reboundmodel1}, except for simulations with different initial angular separations $\Delta \phi_0$. 
    We set $K=2$ in all simulations.
    }
    \label{fig:m1-dphi}
\end{figure}

To study the $\Delta\phi_0$-dependence of our results, we run a suite of simulations with initial $\Delta \phi_0\in [10R_{\rm H}/a_1, 2\pi-10R_{\rm H}/a_1]$, $1/\tau$ between $[0,0.4\Omega_{\rm K}]$, and initial $K=2$.
Figure~\ref{fig:m1-dphi} shows the result, which suggests that
the formation outcome depends weakly on $\Delta \phi_0$.

This result matches the expectation.
For sBHs with initial $\Delta \phi_0\in [10R_{\rm H}/a_1, 2\pi-10R_{\rm H}/a_1]$, their motion remains approximately Keplerian until the mutual separation decrease to a few $R_{\rm H}$ (i.e., when $\Delta \phi(t) \lesssim 10R_{\rm H}/a_1$). 
Therefore, $\Delta\phi_0$ should only affect the timing of the sBH-sBH close encounters but not their outcomes.

\subsubsection{Dependence on $m_{1,2}/M$}
\label{sec:M1_param_m}

\begin{figure}
    \centering
    \includegraphics[width=0.9\columnwidth]{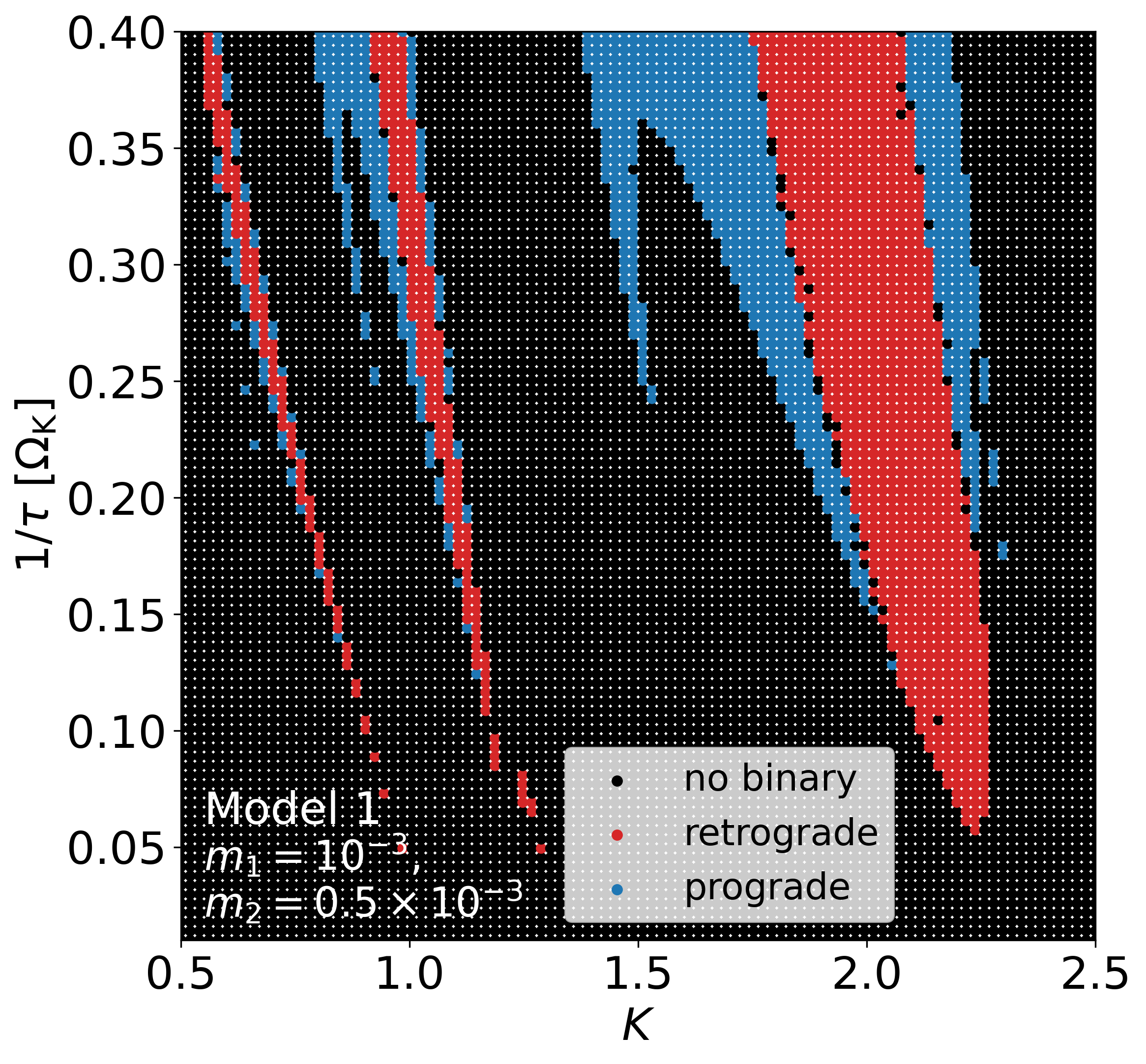}
    \caption{
    Same as Figure~\ref{fig:reboundmodel1}, except for systems with $m_1=10^{-3}M$ and $m_2=0.5\times 10^{-3}M$.
    }
    \label{fig:m1-mass}
\end{figure}

We check the dependence of our results on the masses of the sBHs by repeating the simulations in Section~\ref{sec:M1_param_K} and Figure~\ref{fig:reboundmodel1} with $m_1=10^{-3}M$ and $m_2=0.5\times 10^{-3}M$.
The result is shown in Figure~\ref{fig:m1-mass}

Comparing Figure~\ref{fig:reboundmodel1} and Figure~\ref{fig:m1-mass}, we find that the two are similar: the distribution of formed binaries is only slightly shifted to smaller $K$'s when $m_1$ and $m_2$ increase by three orders of magnitudes.
In Section~\ref{sec:shearing_box}, we demonstrate that the simulation results can be reproduced by a set of equations of motion that are independent of sBH masses (see equations \ref{eq:hilleqx} and \ref{eq:hilleqy}).  Indeed, in the shearing-box local approximation,
the binary formation outcome is independent of the sBH-to-SMBH mass ratio.

\subsection{Multiple Close Encounters}
\label{sec:M1_small_K}

\begin{figure}
    \centering
    \includegraphics[width=1.02\columnwidth]
    {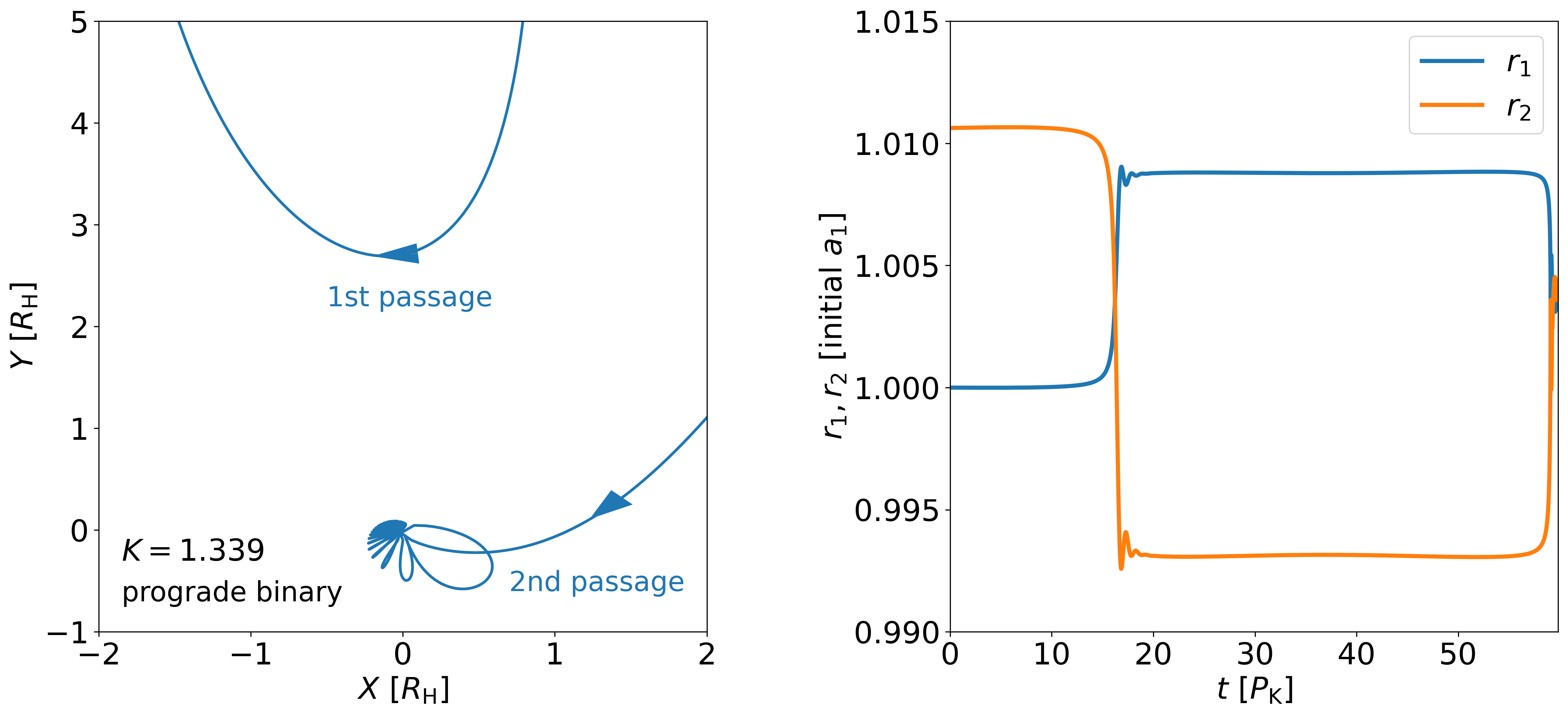}
    \vfill
    \includegraphics[width=1.02\columnwidth]{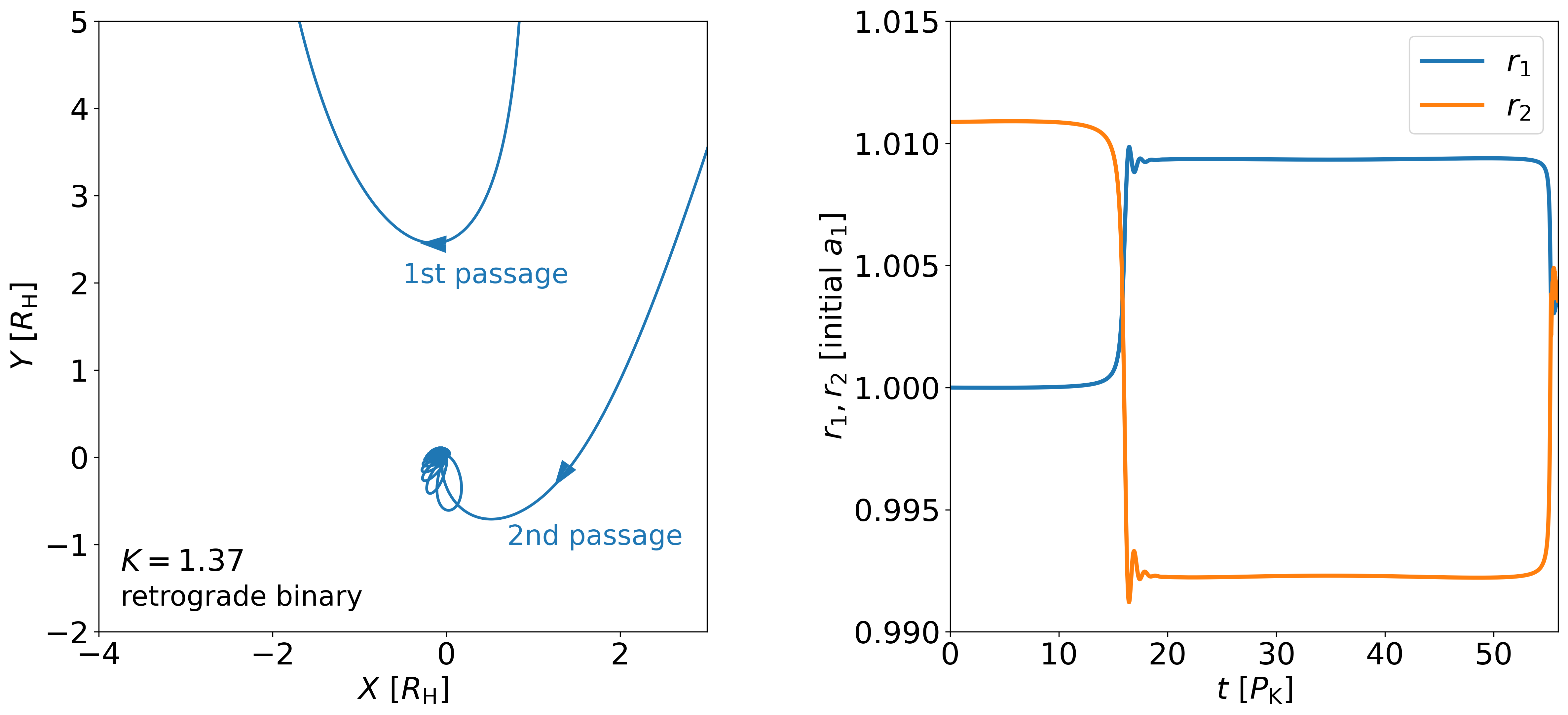}
    \vfill
    \includegraphics[width=1.02\columnwidth]{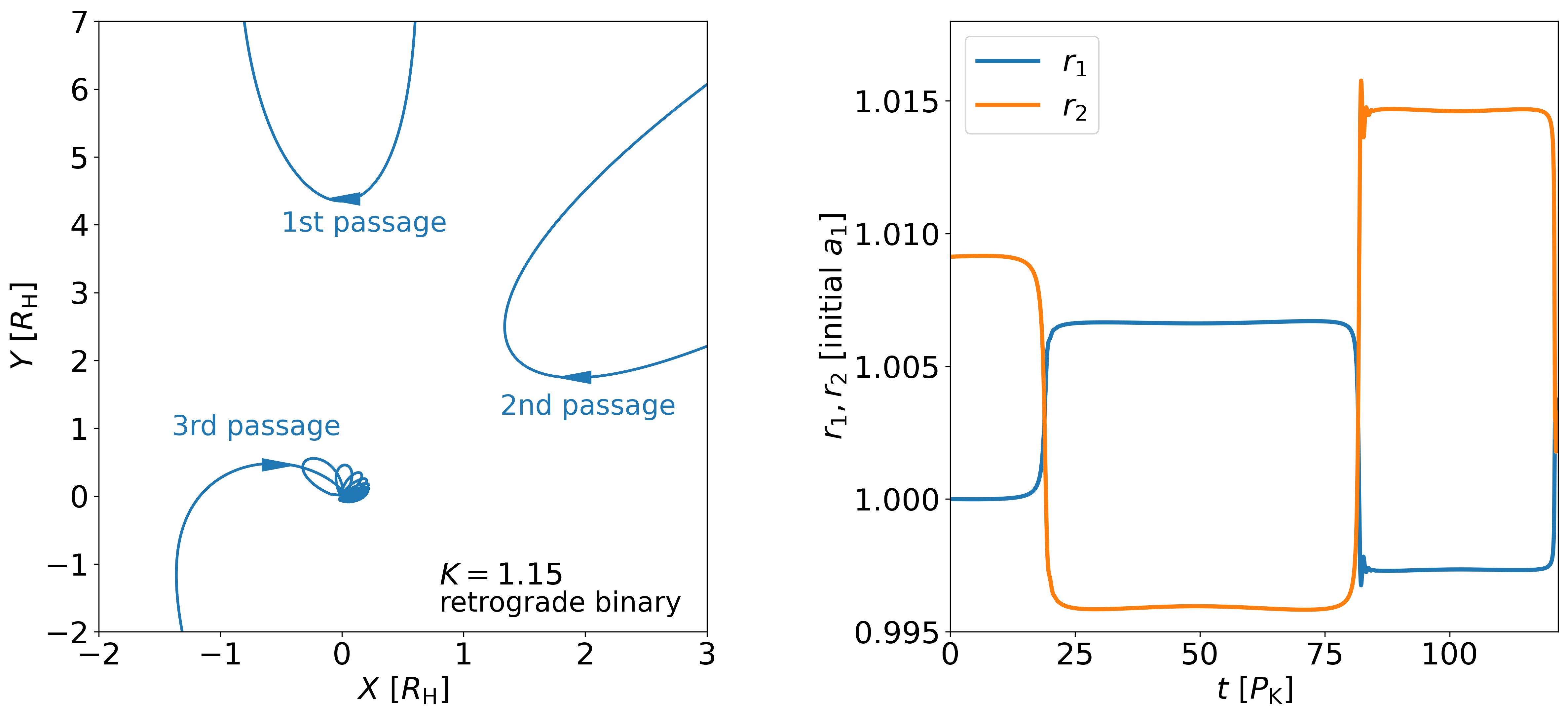}
    \vfill
    \includegraphics[width=1.02\columnwidth]{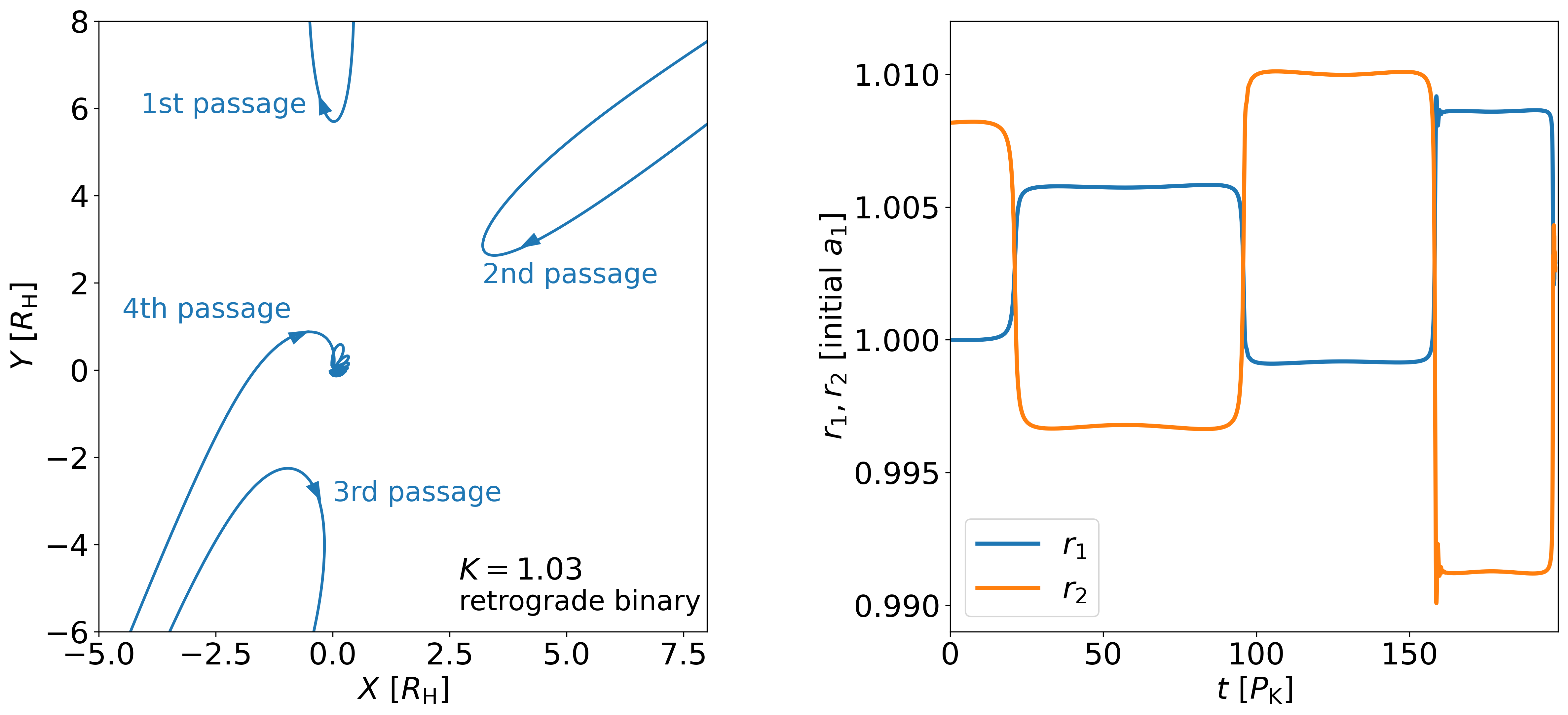}
    \caption{
    Four examples of retrograde/prograde BH binaries that form through consecutive close encounters. The left panels show the relative trajectories in the co-rotating frame of $m_1$ (with $X$ and $Y$ being the radial and azimuthal component of ${\bf r}_2-{\bf r}_1$) during the close encounters, while the right panels show the time evolution of the BHs orbital radii $r_{1,2}$ (distance to the SMBH).
    All cases adopt $1/\tau=0.3\Omega_{1}$, $\Delta \phi_0=\pi/2$, $m_1=10^{-6}M$, $m_2=0.5\times10^{-6}M$, and the values of the initial $K$ as shown in the left panels.
    }
    \label{trajectory_multice}
\end{figure}

In Figure~\ref{fig:reboundmodel1}, most binary formation occurs in systems with initial $K\gtrsim1.7$. When $K\lesssim1.7$, binary formation can only happen inside the narrow finger-like regions around some particular values of $K$.

We find that binaries with initial $K\gtrsim1.7$ are formed during the first BH-BH close encounter, while binaries in the finger-like regions are formed via multiple encounters.
Figure~\ref{trajectory_multice} shows four examples in the latter case.
During their first encounter, the two BHs avoid each other on a horseshoe orbit (see the left panels of Figure~\ref{trajectory_multice}).\footnote{Although we mainly focus on binaries formed during the first sBH-sBH close encounters (see Section~\ref{sec:criteria}), the mutual separation between the sBHs on these horseshoe trajectories during first close encounter remains greater than $>R_{\rm H}$; therefore, they are not ruled out by our criteria of formed binaries. }
The two BHs are then quickly circularized by the gas drag on two new orbits with a large orbital separation (see the right panels of Figure~\ref{trajectory_multice}). 
Therefore, the sBHs will have a larger ``initial'' $K$ during their second close encounter and hence can form a binary just like those cases with a true initial $K\gtrsim1.7$.
This also explains why the substructures of each finger-like region are qualitatively similar to the substructures in the main binary formation region at large $K$.

\section{N-Body Simulation Results: Binary Formation with Gas Friction Model 2}
\label{sec:M2}

In this section, we conduct a series of parameter studies that are similar to those in Section~\ref{sec:M1_param}, but now using the gas dynamical friction (force Model 2). 
Since the results in Section~\ref{sec:M1_param} suggest that the effects of varying $m_{1,2}$ and $\Delta \phi_0$ are weak, we focus on exploring how the formation outcome depends on $K$ and $\tau(x)$.

To make fair comparisons with the results obtained with force Model 1 (Section~\ref{sec:M1_param} and Figure~\ref{fig:reboundmodel1}), the simulations in this Section also use $m_1=10^{-6}M$, $m_2=0.5\times 10^{-6}M$, initial $\Delta \phi_0 = \pi/2$, and initial $K$ between $[1,2.5]$.
The characteristic damping timescale of the GDF $\tau_0$ (see equation~\eqref{eq:tau0}) is chosen such that $1/\tau_{\rm eff} \in [0,0.4\Omega_{\rm K}]$, where
\begin{equation}
    \label{eq:tau_eff}
    \tau_{\rm eff}=\tau\left( x=\frac{\Omega_{\rm K} R_{\rm H}}{c_s}\right).
\end{equation}
When the two sBHs enter their mutual Hill sphere, their relative velocity is of order $\Omega_{\rm K} R_{\rm H}$. Thus, we expect that $\tau_{\rm eff}$ is qualitatively equivalent to $\tau$ in Model 1. 

\begin{figure}
    \centering
    \includegraphics[width=0.9\columnwidth]
        {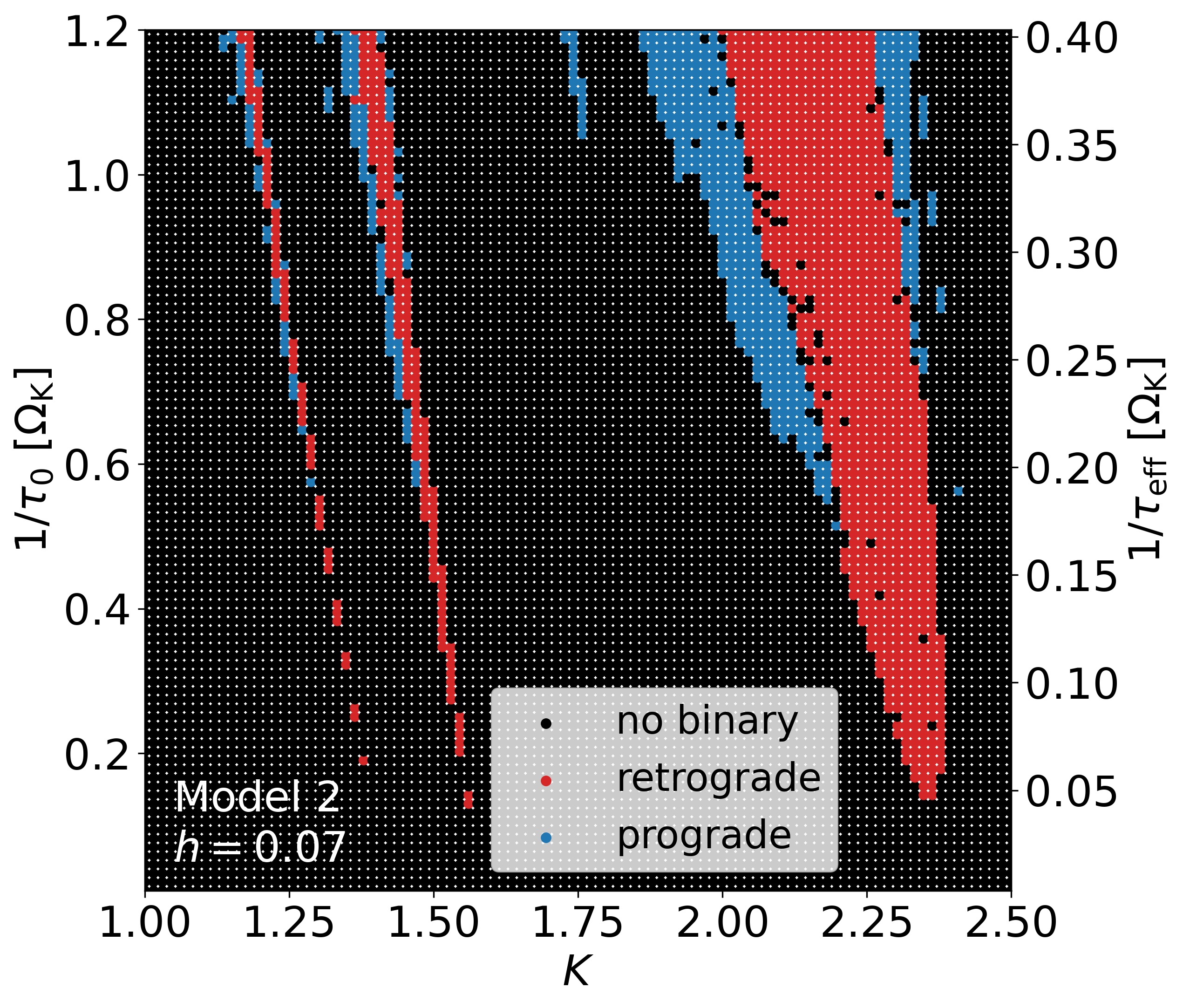}
    \vfill
    \includegraphics[width=0.9\columnwidth]{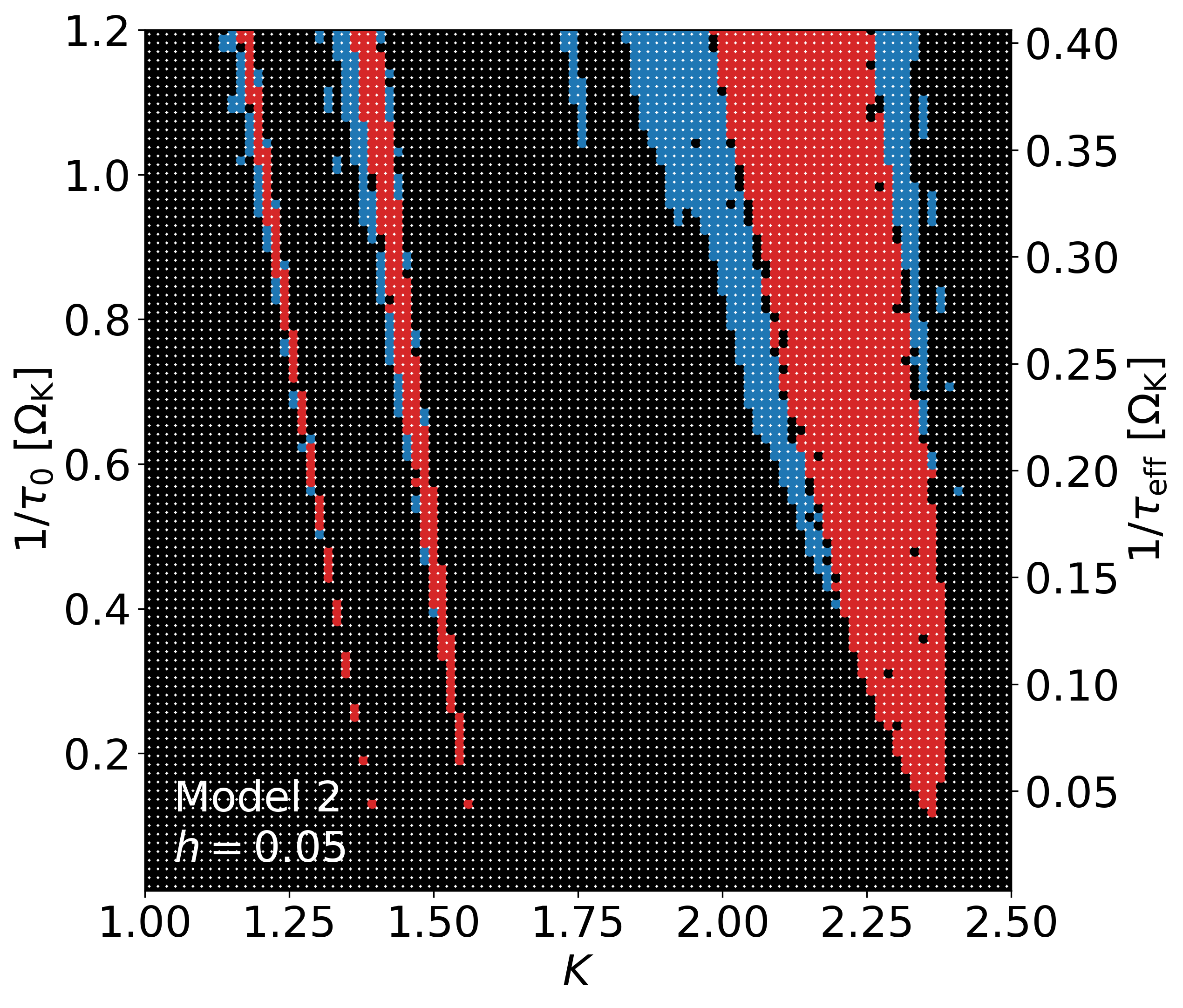}
    \includegraphics[width=0.9\columnwidth]{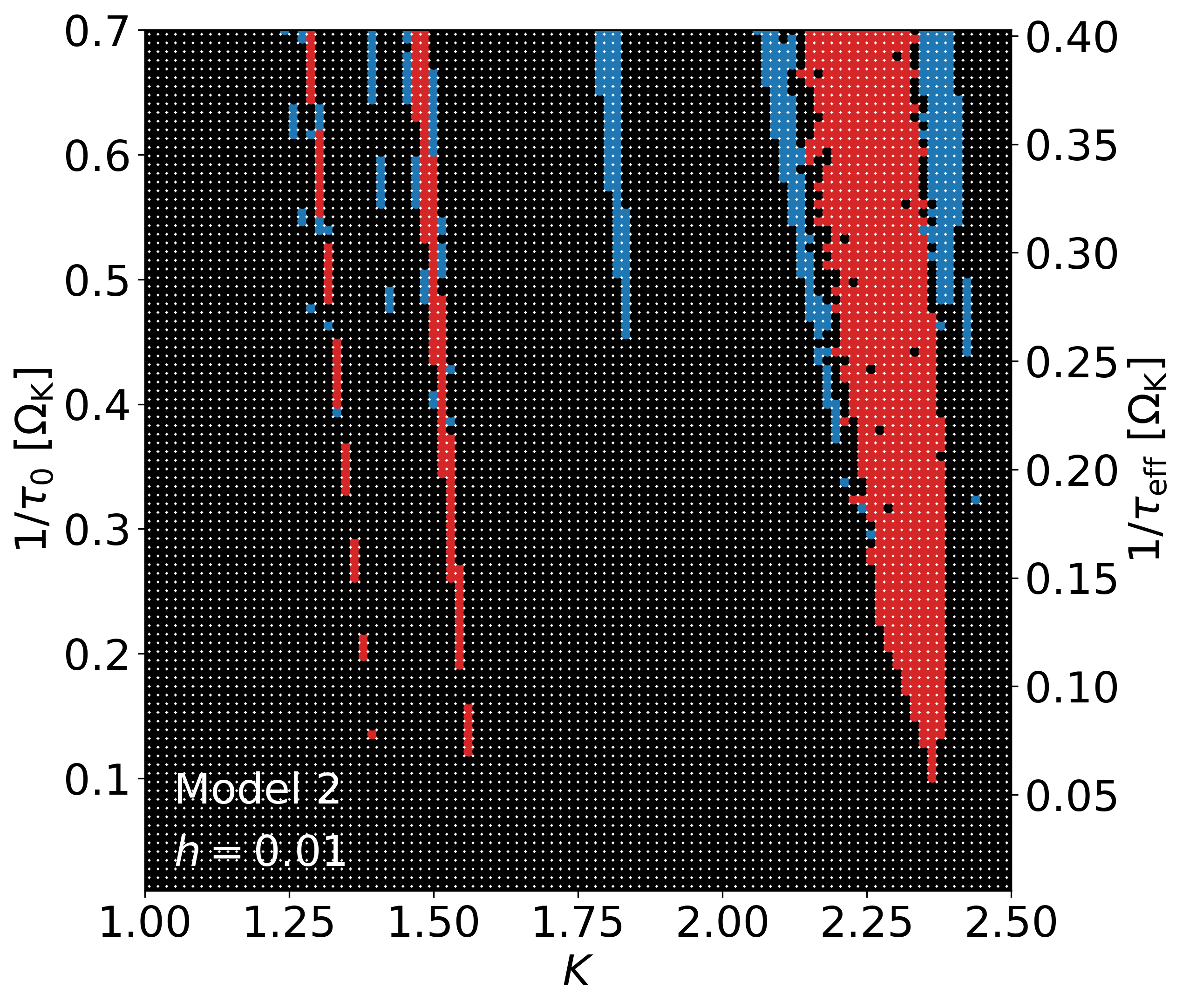}
    \caption{Same as Figure~\ref{fig:reboundmodel1}, except for simulations with the canonical gas dynamical friction (Model 2, equation~\ref{eq:gas-force-2}). Since $\tau_0$ is mass-dependent, the two vertical axes show time scales of the gas drag acting on the inner sBH, $m_1$.
    The second vertical axis on the right indicates the effective dissipation time scale $\tau_{\rm eff}$ (equation \ref{eq:tau_eff}) to compare with the Model 1 result shown in Figure~\ref{fig:reboundmodel1}.
    }
    \label{model2}
\end{figure}

Figure~\ref{model2} shows the results of the parameter studies using Model 2 (GDF). 
Each panel of Figure~\ref{model2} adopts a different $h=c_s/v_{\rm K}$, the disk aspect ratio, but they all demonstrate similar distributions of the formation outcomes.
These results are qualitatively similar to the Model 1 results (Figure~\ref{fig:reboundmodel1}) in terms of the shapes of the formation regimes, the distribution of the prograde and retrograde cases, and the required values of $\tau$ and $\tau_{\rm eff}$ for binary formation.

\begin{figure}
    \centering
    \includegraphics[width=0.9\columnwidth]{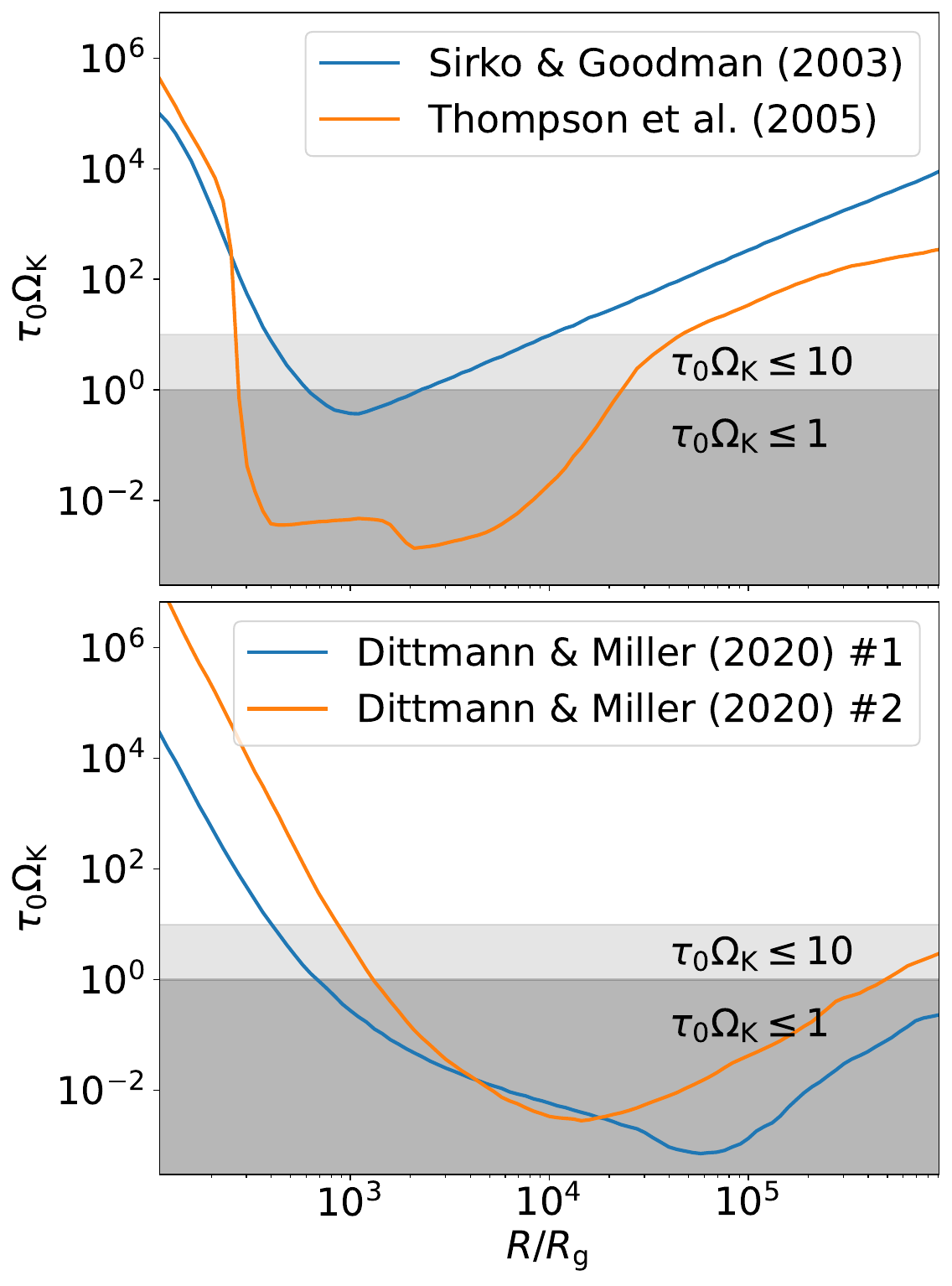}
    \caption{
    Characteristic damping timescale $\tau_0$ for the AGN disk models presented in Figure~2 of \protect\cite{Sirko2003MNRAS}, Figure~6 of \protect\cite{Thompson2005ApJ}, and Figure~1 of \protect\cite{Dittmann2020MNRAS}.
    We extract the $\Sigma$ and $h$ profiles for these models from Figure 1 of \protect\cite{Dempsey2022ApJ}, then calculate $\tau_0$ using equation~\ref{eq:tau0}.
    The upper panel is for SMBHs with $M=10^8M_{\sun}$ and BHs with $m=100M_{\sun}$. 
    The lower panel is for SMBHs with $M=4\times10^6M_{\sun}$ and BHs with $m=4M_{\sun}$.
    }
    \label{fig:disk-tau0-profile}
\end{figure}

In all cases, BH binaries can be formed when $\tau_0^{-1}\gtrsim 0.1\Omega_{\rm K}$ corresponding to $\tau_{\rm eff}^{-1}\gtrsim 0.05\Omega_{\rm K}$; this is similar to the result for model1 (See equation~\eqref{eq:tau_crit}). Higher formation rates require shorter damping timescales.
Figure~\ref{fig:disk-tau0-profile} shows the $\tau_0$ profiles in four different AGN disk models.
The two profiles in the upper panel, originally from \cite{Sirko2003MNRAS} and \cite{Thompson2005ApJ}, are for SMBHs with $M=10^{8}M_{\sun}$, while the lower panel represents the two models from \cite{Dittmann2020MNRAS}, with $M=4\times10^{6}M_{\sun}$.

In the \cite{Sirko2003MNRAS} disk model, the damping timescale is around a few $\Omega_{\rm K}^{-1}$ at radius $R$ (or the semi-major axis of the sBHs) between $[4\times10^2R_{\rm g}, 10^4R_{\rm g}]$, where $R_{\rm g}=GM/c^2$ is the gravitational radius of the SMBH.
Binaries may form within this region. 
However, the formation rate may be low except near $R=10^{3} R_{\rm g}$, where $\tau_0\Omega_{\rm K}\lesssim 1$.
This high-rate location corresponds to the radius where the disk has the smallest aspect ratio $h$ and the largest surface density $\Sigma=2\rho hR$.

The \cite{Thompson2005ApJ} profile has $\tau_0\Omega_{\rm K}\ll1$ in the region $R \in [3\times10^2R_{\rm g}, 10^4R_{\rm g}]$. 
Although the disk in \cite{Thompson2005ApJ} is less dense than in \cite{Sirko2003MNRAS}, it also has an aspect ratio $h\sim10^{-3}$ for $R \in [3\times10^2R_{\rm g}, 10^4R_{\rm g}]$, which is about ten times smaller than in \cite{Sirko2003MNRAS}. 
Therefore, the damping timescale $\tau_0$ can be very short in this thin middle region, leading to higher binary formation efficiency.

\cite{Dittmann2020MNRAS} considers disks around low-mass AGNs at high redshifts. 
Based on their models, $\tau_0\Omega_{\rm K}$ are less than unity from $R\sim10^{3}R_{\rm g}$ to $\sim 10^{6}R_{\rm g}$.
BH binaries can be formed in the middle region of the disk, and also in the outer part of the disk that connects with the star-forming region.
This may allow more progenitor BHs to participate in dynamical encounters and produce a larger population of binaries.

\section{Local simulations with shearing-box model}
\label{sec:shearing_box}

\subsection{Equation of Motion in Shearing-Box}
\label{shearingboxsetup}
In addition to using direct $N$-body simulations, an alternative method to study the sBH-sBH close encounters and binary formation is to use the shearing-box model.
This model applies to the situation when the separation between the BHs is much smaller than the distance to the SMBH (i.e. $|{\bf r}_1-{\bf r}_2|\ll a_1,a_2$), and it solves their relative motion in the co-rotating frame anchored on the center of mass of $m_1$ and $m_2$. 

We assume that the $m_1$-$m_2$ center of mass is circulating around the SMBH at a constant angular velocity $\Omega_{\rm com} \simeq \Omega_{\rm K}\simeq \sqrt{GM/a_1^3}$, which is approximately true when $|{\bf r}_2-{\bf r}_1|$ is small and $m_{1,2}\ll M$.
Keeping terms up to the first order of $|{\bf r}_2-{\bf r}_1|/a_1$, we obtain the shearing-box equations of motion for a BH binary subjected to the gas force:
\begin{align}
    \label{eq:hilleqx}
    \ddot{x}=&3x+2\dot{y}-\frac{3x}{(x^2+y^2)^{3/2}}-\frac{\dot{x}}{\hat{\tau}},\\
    \label{eq:hilleqy}
    \ddot{y}=&-2\dot{x}-\frac{3y}{(x^2+y^2)^{3/2}}-\frac{\dot{y}+3x/2}{\hat{\tau}},
\end{align}
where $x$ and $y$ are the radial and azimuthal components of $({\bf r}_2-{\bf r}_1)/R_{\rm H}$, $\hat{\tau} = \Omega_{\rm com}\tau \simeq \Omega_{\rm K}\tau$ is the dimensionless dissipation time scale, and the dot represents the normalized time-derivative operator ${\rm d}/{\rm d}(\Omega_{\rm K} t)$.
In this section, we consider the gas force Model 1, so that ${\hat \tau}$ is constant.
Equations~\eqref{eq:hilleqx} and~\eqref{eq:hilleqy} are the same as the Hill equations but with the additional terms from the gas force.
A similar set of equations has also been used by \cite{Schlichting2008ApJ02} to study the formation of Kuiper Belt binaries.

The initial condition for the ``SMBH+2sBHs'' systems that we adopt in the previous sections (see Section~\ref{sec:method_ic}) can be implemented in the shearing-box model as
\begin{align}
    \label{eq:Hill_x0}
    x_0&=K,\\
    \label{eq:Hill_y0}
    y_0&=\tan(\Delta \phi_0)a_1/R_{\rm H},\\
    \label{eq:Hill_x0dot}
    \dot{x}_0&=0,\\
    \label{eq:Hill_y0dot}
    \dot{y}_0&=-\frac{3}{2}K
\end{align}
in the limit of $KR_{\rm H}\ll a_1$.

Equations~\eqref{eq:hilleqx}-\eqref{eq:Hill_y0dot} do not depend explicitly on $m_1$, $m_2$, or $M$, which is consistent with $N$-body simulation results in Section~\ref{sec:M1_param_m}.
Nevertheless, we are aware that this shearing-box model is only accurate
when $R_{\rm H}$ is at least two orders of magnitude smaller than $a_1$, i.e. $R_{\rm H}\lesssim 10^{-2}a_1$. 
This is because: (i) the initial velocities of $m_1$ and $m_2$ are Keplerian only when the separation between them is at least a few times of $R_{\rm H}$, i.e $|{\bf r}|\gtrsim 10R_{\rm H}$; 
(ii) the shearing-box approximation requires $|{\bf r}|\ll a_1$. 
To meet both requirements, the BHs need to have $R_{\rm H}/a_1=(m_{12}/3M)^{-1/3}\lesssim 10^{-2}$.
This is equivalent to requiring $m_{12}/M\lesssim 10^{-6}$, which is accordant
with the canonical BH masses in our $N$-body simulations (see sections \ref{sec:M1} and \ref{sec:M2}).

To apply the binary formation criteria (see Section~\ref{sec:criteria}) in the shearing-box model, we compute the relative energy of $m_1$ and $m_2$ as: 
\begin{align}
    \label{eq:Erel_noninertial}
    E_{\rm rel}=&\frac{1}{2}\mu \left|{\bf v}+{\bf \Omega}_{\rm com}\times {\bf r}\right|^2-\frac{Gm_1m_2}{|{\bf r}|}
\end{align}
where ${\bf r}$ and ${\bf v}$ are the relative position and velocity vectors of $m_1$ and $m_2$ in the co-rotating (shearing-box) frame. (Note that ${\bf v}$ is different from ${\bf v}_2-{\bf v}_1$ in equation \ref{eq:Erel_inertial}.) 
Equation~\eqref{eq:Erel_noninertial} can be further simplified by realizing
that ${\bf \Omega}_{\rm com}\simeq \Omega_{\rm K}{\bf \hat z}$ is perpendicular to the plane in which ${\bf r}$ lies.
In dimensionless form, the relative energy is
\begin{align}
    \label{energyshearingbox}
    \epsilon_{\rm rel} 
    & \equiv \frac{E_{\rm rel}}{Gm_1 m_2/R_{\rm H}} \nonumber \\
    & = \frac{1}{6}\left(\dot{x}^2+\dot{y}^2\right)-\frac{1}{\sqrt{x^2+y^2}}+\frac{1}{6}\left(x^2+y^2+2x\dot{y}-2y\dot{x}\right)
\end{align}

\subsection{Parameter Study}

We run a suite of shearing-box calculations with the initial condition
\begin{equation}
    (x,y,\dot{x},\dot{y})_0 = \left(K,10,0,-\frac{3}{2}K\right)
\end{equation}
for $K$ between $[1,2.5]$ and ${\hat \tau}^{-1}$ in the range of $[0,0.4]$.
These ranges for $K$ and ${\hat \tau}$ are to mimic the parameter sweepings in Section~\ref{sec:M1_param_K}.

The value $y_0=10$ corresponds to $\Delta \phi_0=10R_{\rm H}/a_1$, which ensures $10R_{\rm H}\lesssim |{\bf r}_2 - {\bf r}_1|\ll a_1$ initially.
We fix the value of $y_0$ in all of our shearing-box calculations. 
This is because of the weak dependence on $\Delta \phi_0$ seen in our $N$-body results; we expect our shearing-box model results to be approximately independent of $y_0$. 

\begin{figure*}
    \centering
    \includegraphics[width=\textwidth]{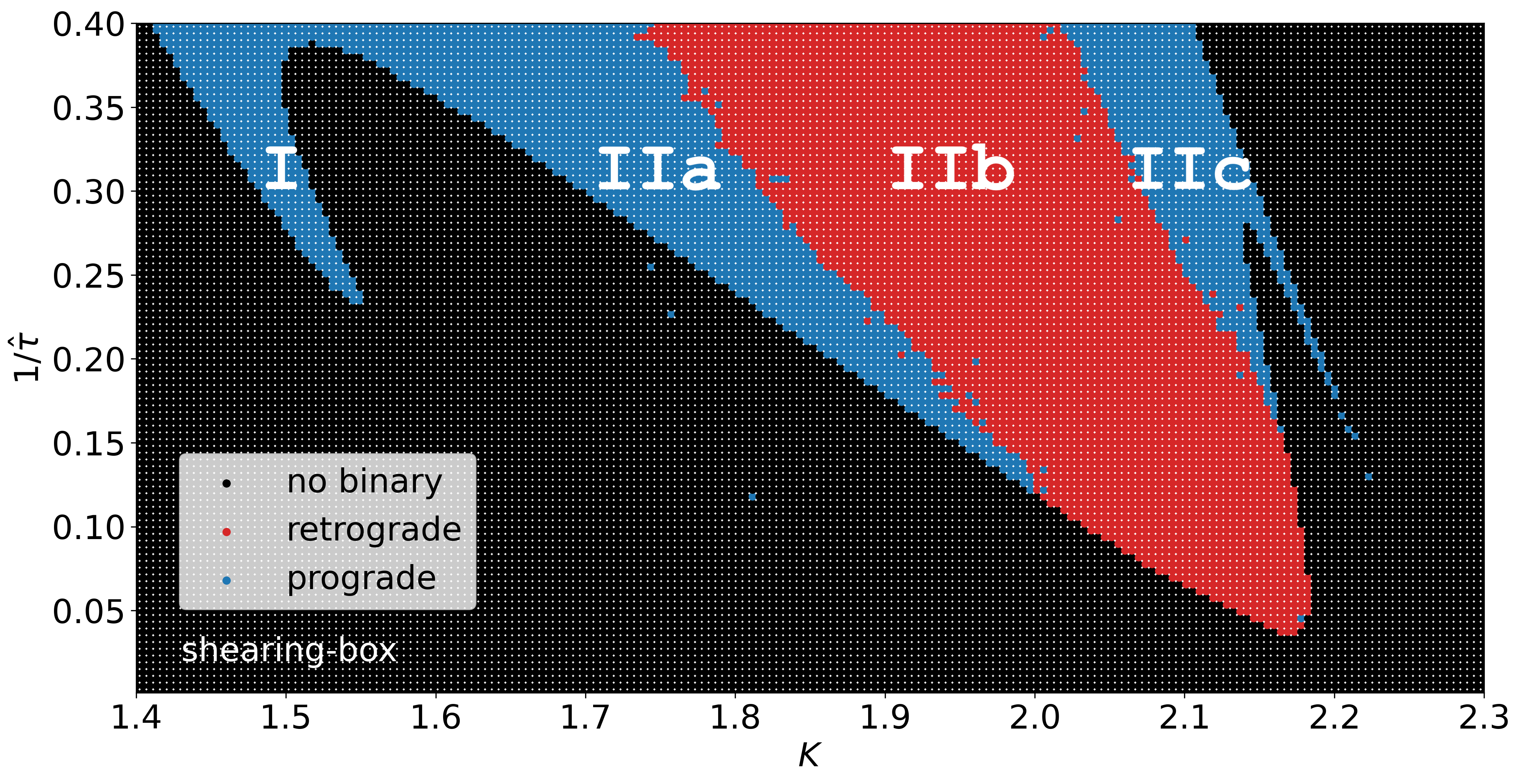}
    \caption{
    Outcomes of our shearing-box model (equations~\ref{eq:hilleqx} and~\ref{eq:hilleqy}) with different initial orbital separations $K$ and frictional damping timescales ${\hat \tau}$.
    All calculations adopt $y_0=10$.
    The black, blue, and red dots represent the systems that do not form binaries, form prograde binaries, and form retrograde binaries, respectively.
    The distribution of the outcomes can be divided into Regions I, IIa, IIb, and IIc.
    BHs evolving from Regions I, IIa, and IIc eventually become prograde binaries, while those from Region IIb form retrograde binaries.}
    \label{shearingbox}
\end{figure*}

Figure~\ref{shearingbox} shows the results.
The blue, red, and black dots represent the formations of prograde binaries, retrograde binaries, and no binary formation, respectively.
Their distribution qualitatively reproduces the results of $N$-body simulations. 
The only difference is that the shearing box does not allow any binary formation at $K\lesssim 1.4$ (i.e. no finger-like regions as seen in Figure~\ref{fig:reboundmodel1}) because the shearing box only describes the first close encounters between the BHs.

\subsection{Binary Orientation: Prograde vs. Retrograde}

\subsubsection{General Description}

The distribution of formed binaries in the $K$-vs-dissipation-rate parameter space can be divided into four regions, corresponding to four types of qualitatively different formation trajectories.
We label the four regions in the Figure~\ref{shearingbox}.
Region I corresponds to the peninsula on the left at $K\approx1.5$.
This region has a lower limit at around $1/{\hat \tau}\approx0.25$ and is filled with prograde binaries.
The rest of the formed binaries are in Region II, which can be further divided into three sub-regions: 
Region IIb refers to the wide band of retrograde binaries at around $K\in[1.8,2.2]$; 
Regions IIa and IIc are the left and the right bands of mostly prograde binaries adjacent to Regions IIb.
When $1/{\hat \tau}$ is between $0.05$ and $0.15$, all binaries inside Region II are retrograde.

We analyze two questions regarding these regions: (i) what determines the binary orientation (prograde-vs-retrograde) in each region, and (ii) why there are only retrograde binaries for $1/{\hat \tau}\lesssim 0.15$.

\begin{figure*}
    \centering
    \includegraphics[height=1.2\columnwidth]{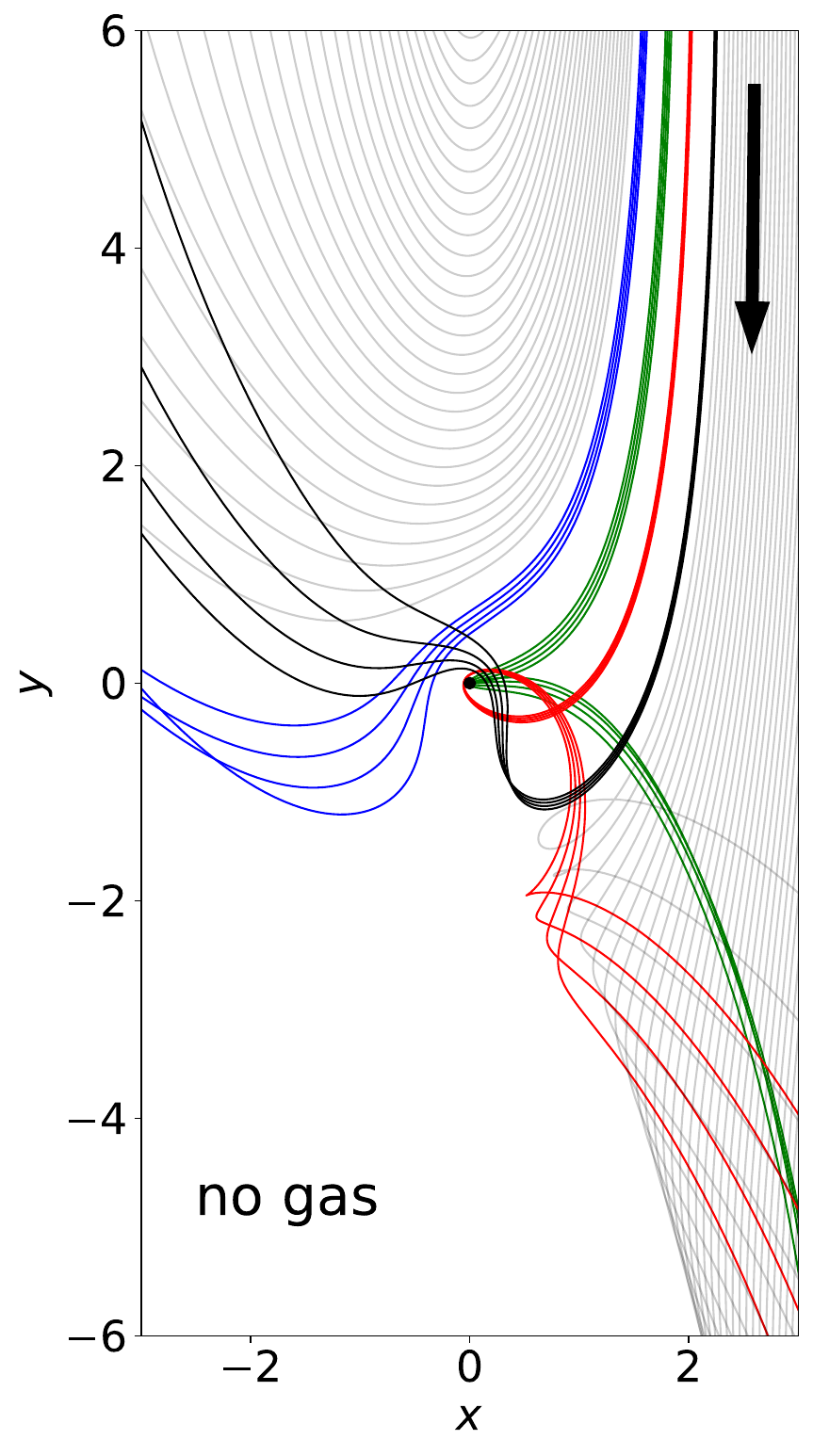} 
    \includegraphics[height=1.2\columnwidth]{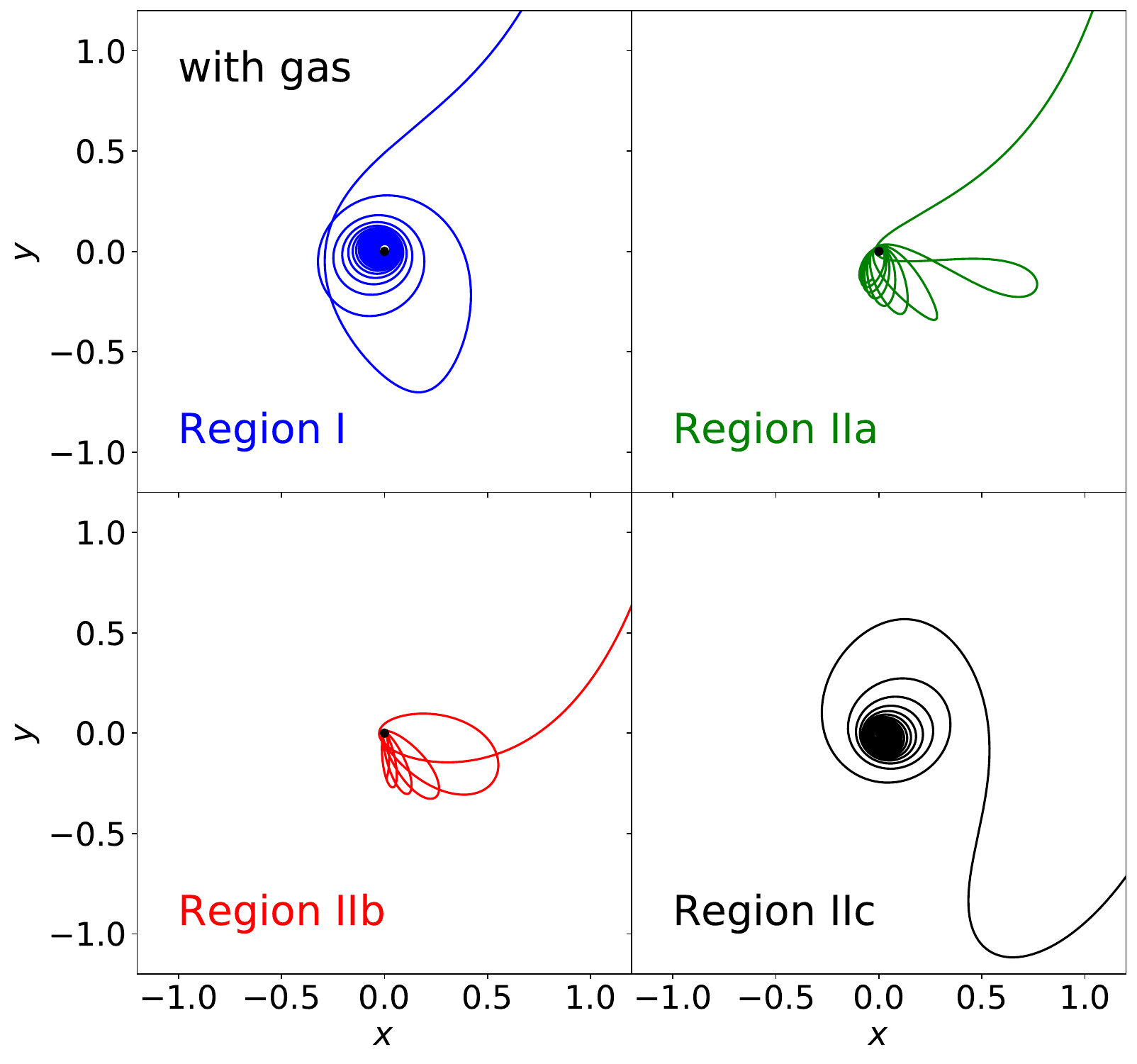}
    \caption{
    \textbf{Left:} Typical trajectories in our shearing-box calculations in the gas-free regime ($1/{\hat \tau}=0$). 
    See the text for the descriptions of the six types of trajectories.
    We do not include boundary cases (e.g., some messy trajectories between the blue and the green curves) for conciseness. 
    \textbf{Right:} Examples from the four binary-forming parameter-space regions depicted in Figure~\ref{shearingbox} (with $1/{\hat \tau}=0.4$). 
    The incoming trajectories (i.e., the pre-capture portion of the curves) are similar to those in the gas-free cases (highlighted in the left panel: blue $\to$ Region I, green $\to$ Region IIa, red $\to$ Region IIb, and black $\to$ Region IIc). 
    }
    \label{fig:region-traj}
\end{figure*}

\subsubsection{Binary Orientation in each Region}
\label{cause_orientation}

Figure~\ref{fig:region-traj} shows some typical trajectories of the BHs in our shearing-box integrations.
In the gas-free regime ($1/{\hat \tau}=0$, see the left panel of Figure~\ref{fig:region-traj}), the dynamical evolution of the BHs is described by the well-known ``satellite-encounter'' problem \citep[e.g.][]{Petit1986Icar}:
\begin{itemize}
    \item For $K\lesssim$1.6 (grey) trajectories, $m_2$ moves in a horseshoe orbit without getting very close to $m_1$.
    \item For  $K$ around 1.65 (blue), $m_2$ is attracted toward $m_1$ in the counter-clockwise direction (prograde), then moves through the left portion of the Hill sphere.
    \item For $K$ around 1.85 (green), $m_2$ is scattered by $m_1$ from the top-right; the close encounter between the two BHs is almost head-on.
    \item For $K$ around 2.0 (red), $m_2$ spirals toward $m_1$ from the right, then completes a full clockwise rotation (retrograde).
    \item For $K$ around 2.27 (black), $m_2$ initially passes $m_1$ from the right, but then makes a U-shape turn toward $m_1$ due to the tidal force and the Coriolis force. Eventually, $m_2$ flies by $m_1$ at a short distance in the counter-clockwise direction (prograde).
    \item For $K\gtrsim$2.35 (grey), $m_2$ passes $m_1$ on an outer orbit without any close encounters.
\end{itemize}
In all cases above (with no friction), $m_2$ eventually leaves $m_1$ and no binaries are formed. 
However, the directions of these close encounters can determine the orientation of the binaries (prograde vs. retrograde) when the gas force is added.

The right panels of Figure~\ref{fig:region-traj} show the dynamical evolution of the BHs in the four parameter-space regions (with $1/{\hat \tau}=0.4$).
The BHs approach each other on trajectories similar to those in the gas-free cases.
As they become tightly bound due to the gas, BHs from Regions I, IIb, and IIc inherit the directions of their encounters as the orientations of the new binaries (i.e., prograde, retrograde, and prograde, respectively).
The BHs from Region IIa experience a near head-on encounter, forming a binary with a very small initial angular momentum.
Before the gas fully stabilizes the binary, $m_2$ first travels to the far-off apocenter on the right. 
It is then deflected by the Coriolis effect to the lower-right quadrant, where the binary relative energy is dissipated by the tidal force.
Hence, as $m_2$ returns, it orbits $m_1$ in the prograde direction as a bound binary.

In summary, the orientation of a binary is determined by the type of its encounter trajectory, which is controlled by the initial orbital spacing $K$.
We also note that a strong gas damping can increase the ``effective $K$'' between two BHs, so the four regions shift to smaller $K$ as $1/{\hat \tau}$ increases.

\subsubsection{Dissipation-Timescale Requirement for Prograde and Retrograde Binaries}

From Figure~\ref{shearingbox}, we see that all binaries are retrograde when $1/{\hat \tau}\lesssim 0.15$.
Similar results have been reported by \cite{Schlichting2008ApJ10} for the formation of Kuiper Belt binaries due to (particle) dynamical frictions. 
\footnote{\cite{Schlichting2008ApJ10} have only considered values of $1/{\hat \tau}$ ranging from 0.0004 to 0.1, so their calculations do not produce any prograde Kuiper Belt binaries (for the $L^2$s mechanism).}
\cite{Schlichting2008ApJ10} suggest that this is because the retrograde binaries are more stable than the prograde ones.
One possible explanation for the stability difference is the Coriolis force:
during the binary formation process, the Coriolis force always points outward for prograde binaries and inward for retrograde binaries, and this tends to destabilize and stabilize their orbits, respectively \cite[see, e.g.,][]{Hamilton1991Icar,Hamilton1997Icar}.
This difference in the stability could imply that, for BH pairs that are trying to form binaries, those undergoing retrograde encounters are more likely to succeed.

To gain some insight on prograde \& retrograde binary formation, we rerun our shearing-box calculations for $(K,1/{\hat \tau})\in[2,2.2]\times[0.05,0.15]$, where all bound binaries are retrograde; 
however, at some point during the binary formation process, we reverse the relative velocity of the BHs from $(\dot{x},\dot{y})$ to $(-\dot{x},-\dot{y})$ and check whether they can form prograde binaries instead.
Our sample contains $822$ systems that would form stable retrograde binaries without the velocity change. 
When we reverse the relative velocity of the BHs during their first pericenter passage, we find that $407$ systems remain bound as wider retrograde binaries and the rest become unbound.
If we wait for the BHs to pass the pericenter completely and then reverse their velocity when their separation $\sqrt{x^2+y^2}$ becomes less than 0.25, $6$ out of $822$ systems can form stable prograde binaries, and the rest become unbound.
If we reverse the velocity at the first apocenter, all systems fall apart.
This experiment shows that the prograde binaries are indeed less stable than the retrograde ones. 
Hence, \textit{prograde binaries require stronger gas friction to form, while retrograde ones can still form when the friction is weaker}.

\section{GW-Emission-assisted Formation of Binaries}
\label{sec:GW-binary}

In previous sections, we have considered binary formation when the only dissipative process is the gas drag.
However, when the sBHs reach a sufficiently small separation, GW emission becomes important.
Two encountering sBHs can form a tight binary via GW bremsstrahlung when their pericenter distance $r_{\rm p}$ becomes less than a critical value $r_{\rm GW}$ \citep{LJR2022ApJ}, 
\begin{equation}
\frac{r_{\rm GW}}{R_{\rm H}}\simeq 10^{-4} \eta^{-2/7}\left( \frac{4\mu}{m_{12}}\right)^{2/7}\left( \frac{10^6 m_{12}}{M}\right)^{10/21}\left(\frac{a_1}{100R_g} \right)^{-5/7}
\label{eq:rGW}\end{equation}
where $\eta$ is a constant of order unity. 
For our fiducial sBHs at $a_1 \sim 10^3R_g$, at which the gas-assisted binary formation is efficient (see Figure~\ref{fig:disk-tau0-profile}), $r_{\rm GW}\sim2\times 10^{-5} R_{\rm H}$.

\subsection{Method}
We re-run our fiducial shearing-box calculations incorporating the GW-assisted binary formation:
while we simulate encountering sBHs with equations~\eqref{eq:hilleqx} and~\eqref{eq:hilleqy}, we immediately consider them as a bound binary once their mutual separation becomes smaller than $r_{\rm GW}=10^{-5}R_g$. 
For simulations with $({\bf r}_2-{\bf r}_1)/R_{\rm H}>r_{\rm GW}$ at all time, we use the same binary-formation criteria as in the fiducial shearing-box calculation and the $N$-body studies (see Section~\ref{sec:criteria}).

\subsection{Results}

\begin{figure*}
    \centering
    \includegraphics[width=\textwidth]{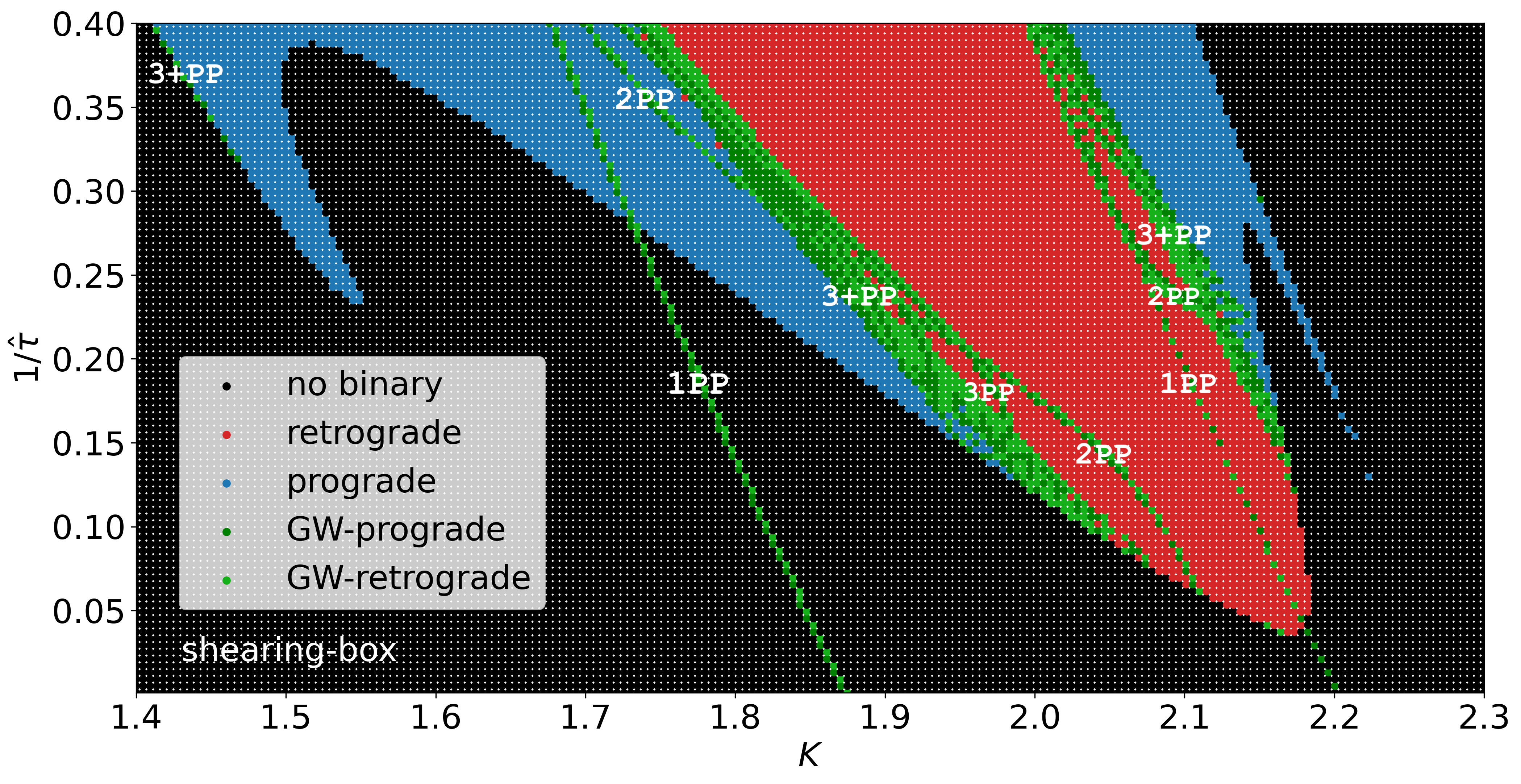}
    \caption{Same as Figure~\ref{shearingbox} except that here we include binary formation by GW emission in addition to the gas drag.
    The blue and red dots represent binaries formed without triggering GW emission, i.e. the only dissipative process being gas drag.
    The deep green and light green dots represent prograde and retrograde binaries formed due to GW emission, respectively.
    The labels ending with ``PP'' on the green dots indicate during which episode of pericenter passages the system reaches $|{\bf r}_1-{\bf r}_2|<r_{\rm GW}=2\times 10^{-5}R_{\rm H}$ and form a bound binary due to GW emission.
    }
     \label{fig:shearingbox_GW}
\end{figure*}

Figure~\ref{fig:shearingbox_GW} shows the shearing-box calculation results when we include binary formation by GW emission.
In addition to the blue, red, and black dots defined in the same way as in Figure~\ref{shearingbox}, the deep green and light green dots represent prograde and retrograde binaries formed due to GW emission, respectively.
Most of these cases exhibit highly chaotic orbital trajectories, and the sBHs can sometimes undergo multiple pericenter passages before they enter the GW-capture regime.

In the absence of gas friction ($1/{\hat \tau}=0$), binary formation by GW emission occurs only when $K$ is around two narrow ranges centered at $K\simeq 1.87$ and $2.2$, corresponding to the two cases where the two sBHs experience close encounters with nearly zero relative angular momentum (see the left panel of Figure~\ref{fig:region-traj}). 
\cite{Boekholt2023MNRAS} have shown that within these narrow ranges, the character of the trajectory depends sensitively on the precise value of $K$, and exhibits a fractal-like structure.

In Figure~\ref{fig:shearingbox_GW}, the cases where the sBHs enter the GW-dominated regime ($r_{\rm p}\lesssim 10^{-5}R_{\rm H}$) during their first pericenter passage are represented by two thin green lines passing through $(K,1/{\hat \tau})\approx(1.9,0)$ and $(2.2,0)$.
However, most GW-induced binary formation occurs during the second or later pericenter passages. 
These binaries are distributed at the boundaries between the gas-dominated-prograde and retrograde regions. 
One can think of them as binaries that are initially ``pre-captured'' by the gas drag (like their prograde and retrograde neighbors) and then enter the GW regime due to their low angular momentum.

\section{Binary formation rate}
\label{sec:FR}

In the previous sections, we have obtained the probability of binary formations as a function of $K$ and the friction timescale $\tau$. 
These results can be applied to a ``realistic'' situation where many sBHs populate the AGN disk, and undergo two-body close-encounters. 
Let $n$ be the number density (per unit area) of sBHS in a certain local region of the AGN disk. 
The binary formation rate per unit area is
\begin{equation}
\label{eq:FR1}
    {\cal R} = \frac{1}{2} n\int_{-\infty}^\infty F(K, \tau) n v_{\rm shear} R_{\rm H} dK,
\end{equation}
where $R_{\rm H}$ is the Hill radius, $v_{\rm shear} \simeq(3/2)\Omega_{\rm K} R_{\rm H} |K|$ is the shear velocity (with $\Omega_{\rm K}$ as the initial Keplerian angular velocity around the SMBH, see equation \ref{eq:Omega_Kep}). 
\footnote{We assume all sBHs initially move on circular orbits in the AGN disk plane. For simplicity, we also neglect the dependence of the result on $m_1/m_2$, effectively assuming all sBHs have similar masses.}
In equation \eqref{eq:FR1}, $F(K,\tau)$ is the ``formation outcome'' function, 
\begin{equation}
    \label{formationoutcome}
    F(K, \tau)=
    \begin{cases}
    1,\ \text{if the BHs form a binary at $(K,\tau)$},\\
    0,\ \text{if the BHs do not form a binary $(K,\tau)$},\\
    \end{cases}
\end{equation}
based on the results shown in Figure~\ref{fig:reboundmodel1},~\ref{fig:m1-mass},~\ref{model2}, or~\ref{shearingbox}. 
Binary formation only happens for values of $|K|$ ranging from 1 to 2.5.
Within this range, we may assume that $n$ is a constant. Since $F(K,\tau)=F(-K,\tau)$, we can rewrite equation \eqref{eq:FR1} as
\begin{equation}
\label{eq:FR2}
    {\cal R} = \frac{3}{2}n^2\Omega_{\rm K} R_{\rm H}^2\int_{1}^{2.5} F(K, \tau)K dK.
\end{equation}
Similar equations can be used to estimate the rates of forming prograde and retrograde binaries.

\begin{figure*}
    \centering
    \includegraphics[height=0.9\columnwidth]
    {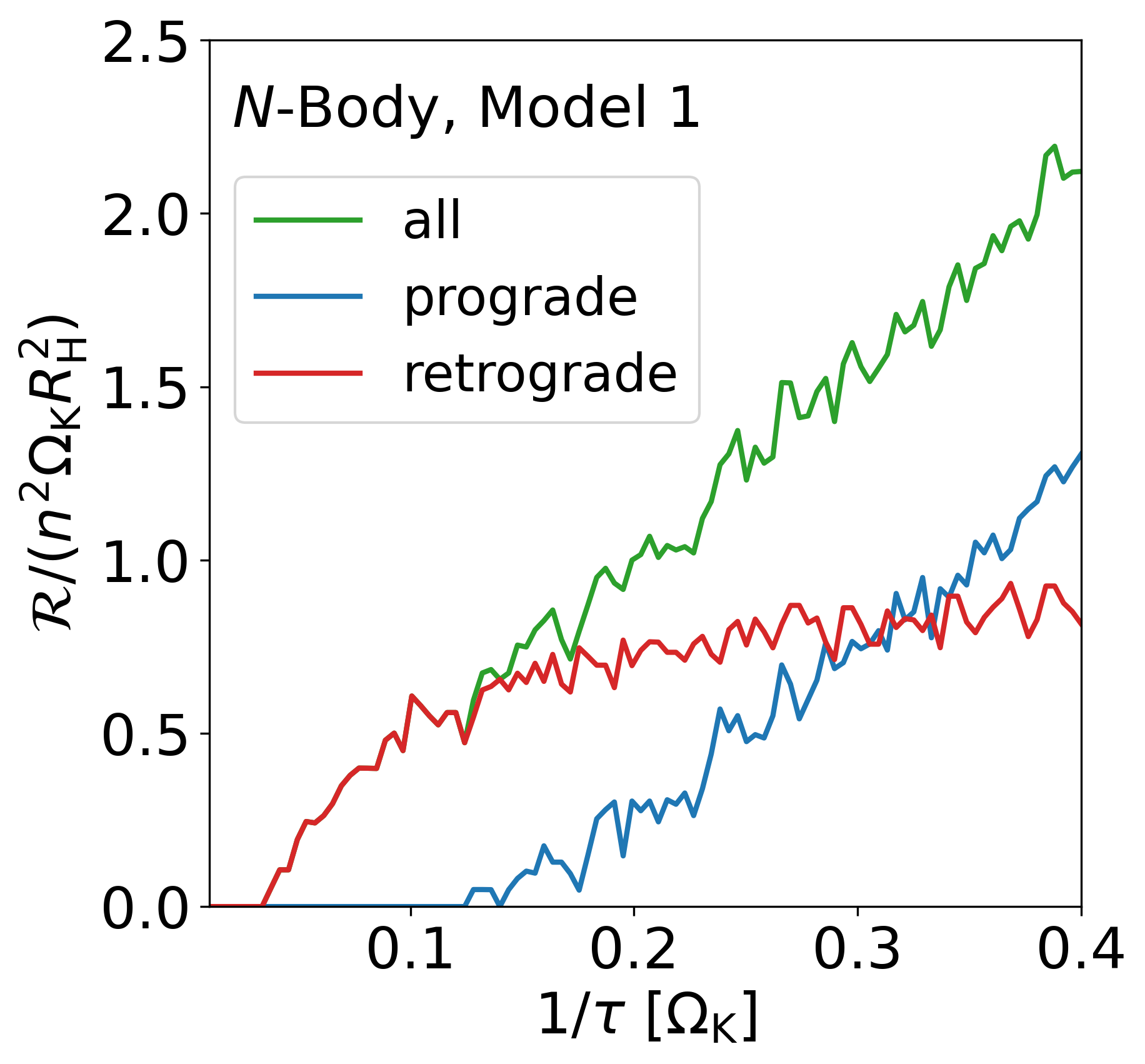}
    \hfill
    \includegraphics[height=0.9\columnwidth]{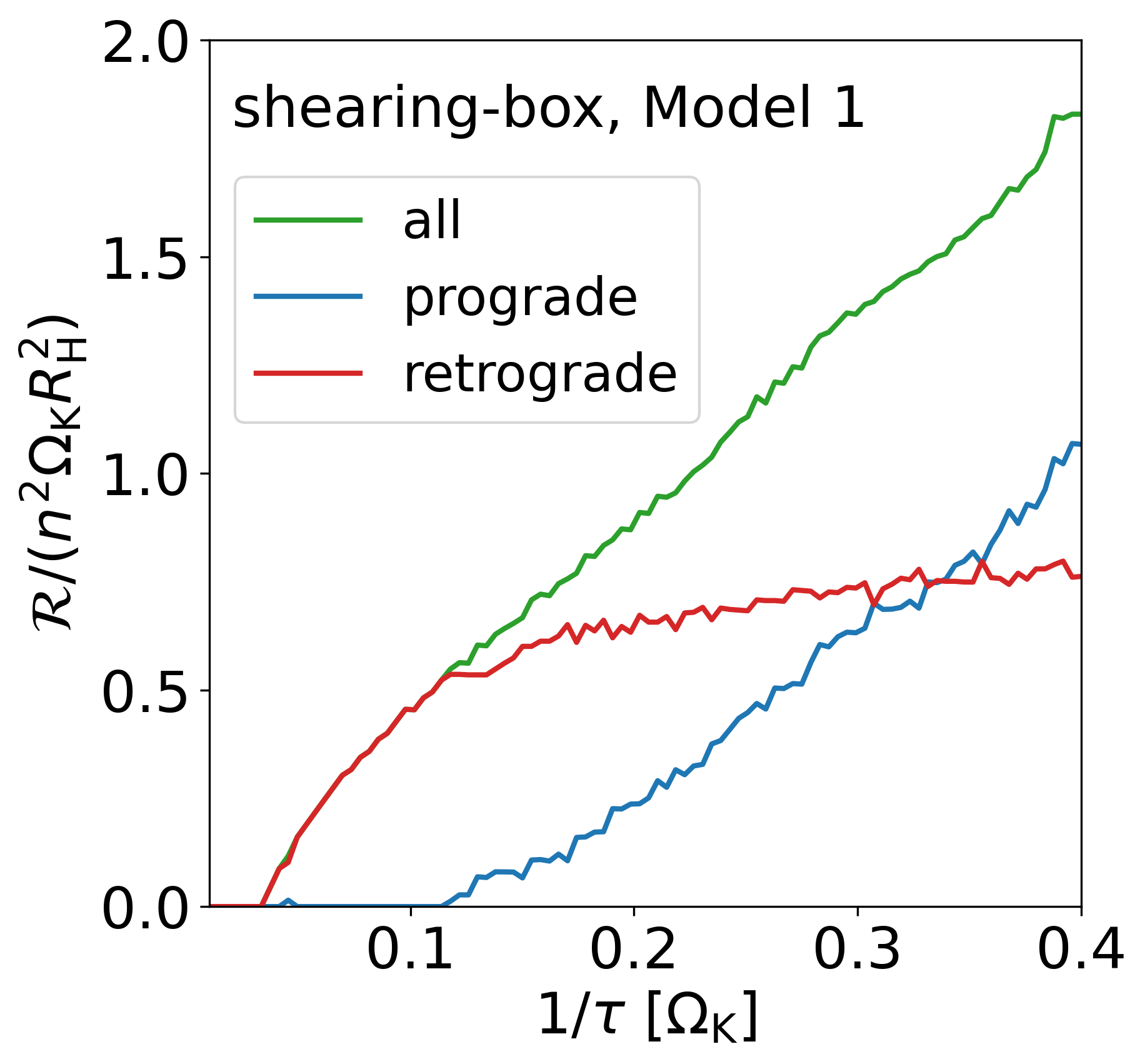}
    \caption{
    {\bf Left:} Prograde (blue), retrograde (red), and total (green) binary formations calculated based on equation~\eqref{eq:FR2} and the Model 1 results (linear gas friction, Figure~\ref{fig:reboundmodel1}). {\bf Right:} Same as the left panel, except for using the shearing-box calculation.
    Note that the jaggedness of the curves (especially in the left panel) results from the finite number of grid points in the ($K$-$\tau^{-1}$) parameter space of our calculations and from the sensitive dependence of binary formation on the value of $K$, especially in the borderlines region (see, e.g. Figure~\ref{fig:reboundmodel1}).
    }
    \label{FR_model1}
\end{figure*}

\begin{figure*}
    \centering
    \includegraphics[width=\textwidth]{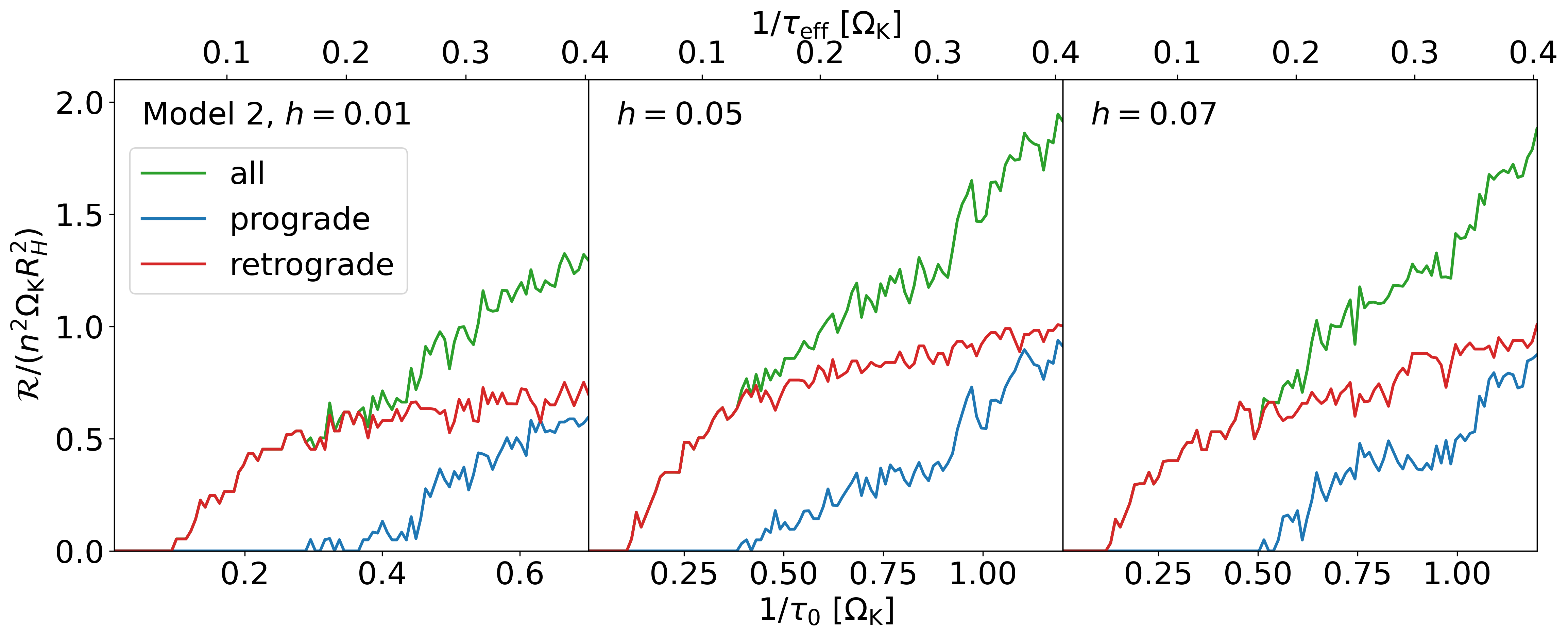}
    \caption{
    Same as Figure~\ref{FR_model1}, except for Model 2 results (gas dynamical friction, Figure~\ref{model2}).
    }
    \label{FR_model2}
\end{figure*}

The left panel of Figure~\ref{FR_model1} shows the formation rate estimated by equation~\eqref{eq:FR2} using the $F(K,\tau)$ from Figure~\ref{fig:reboundmodel1} (i.e., $N$-body Model 1), while the right panel shows the formation rate estimated from Figure~\ref{shearingbox} (i.e., shearing-box results). 
The two panels are similar since the shearing-box results are the same as the Model 1 $N$-body results except for the binaries formed through multiple close encounters.
Retrograde binaries begin to form at around $1/\tau \simeq 0.03\Omega_{K}$ (see equation \eqref{eq:tau_crit}),
Their formation rate reaches around $0.5n^2 R_{\rm H}^2\Omega_{K}$ at $1/\tau=0.1\Omega_{K}$ saturates at around $n^2 R_{\rm H}^2\Omega_{K}$ for larger $1/\tau$.
Prograde binaries appear at $1/\tau\gtrsim0.13\Omega_{K}$; their formation rate grows as the gas force increases, and surpasses the retrograde rate at around $1/\tau\gtrsim0.32\Omega_{K}$.
The total formation rate is roughly a linear function of $1/\tau$.
We also calculate these rates based on Figure~\ref{model2} (gas force Model 2), and show the results in Figure~\ref{FR_model2}.
All these results are qualitatively similar, except that the $h=0.01$ case has a slightly lower overall formation rate than the others. 

We noted that, as depicted in Figures \ref{FR_model1} and \ref{FR_model2}, the binary formation rate becomes negligibly small when the gas friction rate $1/\tau$ drops below the critical value$\simeq 0.03\Omega_{\rm K}$ (which depends on the gas drag models). 
As discussed in Section \ref{sec:GW-binary}, the rate is not zero even in the limit of $1/\tau\rightarrow 0$ since binary formation can be facilitated by GW emission. We can estimate the ``gas-free'' binary formation rate as follows. When two sBHs have the initial orbital separation $K$ centered around the special value $K_0\simeq 1.87$ or $K_0\simeq 2.2$, they enter the Hill sphere and approach each other with nearly zero relative angular momentum. Since the typical velocity inside the Hill sphere is of order $v_{\rm H}\simeq \Omega_K R_{\rm H}\sim \sqrt{Gm_{12}/R_{\rm H}}$, we can estimate the effective width $\Delta K$ (around $K_0$) within which the two sBHs enter the GW-dominated regime as 
\begin{equation}
    (R_{\rm H} \Delta K) v_{\rm H}\sim \sqrt{2 Gm_{12}r_{\rm GW}}
\end{equation}
which gives $\Delta K\sim (r_{\rm GW}/R_{\rm H})^{1/2}$. Thus, the dimensionless binary formation rate in the gas-free limit is 
\begin{equation}
    \frac{R}{n^2\Omega_K R_{\rm H}^2}\sim \Delta K \sim \left( \frac{r_{\rm GW}}{R_{\rm H}} \right)^{1/2}=3\times 10^{-3} \left( \frac{r_{\rm GW}}{10^{-5}R_{\rm H}} \right)^{1/2}
\end{equation}
This is consistent with the scaling result obtained by \cite{LJR2022ApJ}.

\section{Summary and Discussion}
\label{sec:summary}

\subsection{Summary}

In this paper, we have studied the formation of stellar-mass black hole (sBH) binaries in AGN disks due to the frictional forces from the gas.
We have focused on the ``SMBH + 2sBH'' systems, where the sBHs are initialized on circular and co-planar orbits separated by $K$ times the Hill radius $R_{\rm H}$.
When $K$ is small ($\lesssim 2.5$), the sBHs quickly experience a close encounter, during which they may capture each other and form a tightly bound binary due to the energy dissipation caused by gas. The key results of this paper are summarised as follows.

(i) Using $N$-body simulations with a linear frictional force, we systematically investigate the conditions of binary formation and the orientation distributions (prograde vs. retrograde) of the newly formed binaries (Section~\ref{sec:M1}).
We show that the formation probability depends strongly on the initial orbital separation $K$ and the frictional damping timescale $\tau$ (Figure~\ref{fig:reboundmodel1}). 
In particular, retrograde binaries begin to form when $\tau^{-1}\gtrsim 0.03\Omega_{\rm K}$ and prograde binaries begin to form when $\tau^{-1}\gtrsim 0.1\Omega_{\rm K}$.
The results depend weakly on the initial orbital phase difference $\Delta \phi_0$ (Figure~\ref{fig:m1-dphi}), and are qualitatively the same for sBHs with of different masses $m_{1,2}/M$ (Figure~\ref{fig:m1-mass}). These $N$-body results can be reproduced by shearing-box calculations (Section~\ref{sec:shearing_box}), except that $N$-body 
calculations allow for binary formation during two or more close encounters.  
Multiple features of the distribution of the binary orientations in the $K-\tau^{-1}$ parameter space are identified
(Regions I, IIa, IIb, and IIc in Fig.~\ref{shearingbox});
these are associated with four groups of qualitatively different trajectories in the gas-free regime (see Fig.~\ref{fig:region-traj}).

(ii) We rerun some simulations using the gas-dynamical friction (GDF) forces (Section~\ref{sec:M2}).
The results are similar for the two gas-force models when $\tau$ (of the linear friction, see equation~\ref{eq:gas-force-1}) and $\tau_{\rm eff}$ (of GDF, see equation~\ref{eq:tau_eff}) are similar.
Based on equation~\eqref{eq:tau0} and the AGN disk models in the literature, we estimate the possible disk locations where BBHs may form (Figure~\ref{fig:disk-tau0-profile}).

(iii) We also rerun our shearing-box calculations that include the 
possibility of binary formation via GW emission at very small binary 
separation ($r\lesssim r_{\rm GW}\ll R_{\rm H}$; 
see Eq.~\ref{eq:rGW}). 
These GW-assisted binary formations occupy
specific domains in the $K-\tau^{-1}$ parameter space (see Fig.~\ref{fig:shearingbox_GW}), depending on the number of pericenter passages the sBHs undergo before GW emission dominates that leads to binary capture. 
The binary orientations can be either prograde or retrograde.

(iv) Based on our numerical results for the formation probability as a function of $K$ and $\tau$, we obtain the rate of binary formation in AGN disks as a function of the sBH number density $n$ and the gas damping timescale $\tau$ (or $\tau_0$) (Section~\ref{sec:FR}).
The results show that the binary formation rate increases approximately linearly with $1/\tau$ (Figures~\ref{FR_model1} and~\ref{FR_model2}) when $1/\tau$ is above a threshold ($\simeq 0.03\Omega_{\rm K}$); below this threshold, GW-assisted capture dominates. The ratio between the prograde and retrograde binaries depends on $\tau$ or $\tau_0$: 
when the gas friction is weak (small $1/\tau$), only retrograde binaries may be assembled; when the gas friction is strong, prograde binaries may become equally likely as retrograde binaries.

\subsection{Discussion}

We mention several caveats of our study. First, all our calculations assume that the two sBHs initially have circular orbits and zero inclination relative to the disk (so that their motion remains co-planar). In reality, the sBHs may have a small mutual inclination 
and orbital eccentricities, and these may affect some aspects of our results (see \cite{rom2023formation}).
Our preliminary studies suggest that an inclination $\sim 1^\circ$ has a negligible effect on the gas-dominated formation of binaries. 
This is because, to form a binary, the gas damping timescale $\tau$ needs to be less than about $5P_{\rm K}$, which is considerably shorter than the synodic period between the sBHs.
This implies that the gas drag is able to damp most of the out-of-plane motion of the sBHs before the BH-BH encounter, so that the formation outcomes are the same as in the co-planar cases.
However, binary formation through GW emission may be sensitive to the residual inclinations, because $r_{\rm GW}/R_{H}$ is very small.
Future studies need to examine this effect in detail \citep[see][for the gas-free scenario]{LJR2022ApJ}.

Second, we have only studied two-body encounters in this paper.
Multiple sBHs may reside in the same region of the AGN disk (e.g. due to migration trap). Interactions between three or more sBHs may lead to new features of binary BH formation.

Finally, the most important caveat of our study is our treatment of the gas effect. 
Although the two gas-friction models examined in this paper yield qualitatively similar results in terms of the binary formation outcomes in the $K-\tau^{-1}$ parameter space, both models are highly simplified compared to reality. 
Hydrodynamical simulations show that when two sBHs undergo close encounters, the dominant interaction involves the collision of the mini-disks around each BH and the post-collision drag \citep{LJR2023ApJL,Whitehead2023arXiv}.
More studies using hydrodynamics simulations are needed to obtain a more definitive understanding of the gas effects on binary BH formation in AGN disks.

\acknowledgments
This work has been supported in part by the NSF grant AST2107796
and by Cornell University.
Jiaru Li was supported in part by a CIERA Postdoctoral Fellowship.

\software{
IAS15 \citep{Rein2015MNRAS},
Matplotlib \citep{Hunter2007}, 
NumPy \citep{Walt2011},
Rebound \citep{Rein2012aap},
SciPy \citep{Virtanen2020}
}


\end{document}